\documentclass[sigconf,nonacm]{acmart}

\AtBeginDocument{%
  }

\usepackage{algorithm, algorithmic,multirow,url,xcolor}
\usepackage{amsfonts,textcomp,verbatim}
\usepackage{graphicx}
\usepackage{framed}
\usepackage{fancybox}
\usepackage{balance}
\usepackage{hyperref}

\newcommand{\sysName}{OCR-APT}
\newcommand{\gnnModel}{OCRGCN}
\newcommand{\detector}{subgraph anomaly detector}
\newcommand{\investigator}{attack investigator}

\def\shorten{\looseness=-1}
\newcommand{\RDFTYPE}[3]{\ensuremath{{\langle}\texttt{#1},} \texttt{#2}, \ensuremath{\texttt{#3}{\rangle}}}

\renewcommand\footnotetextcopyrightpermission[1]{} 
\setcopyright{none}

\begin{document}

\title[OCR-APT: Reconstructing APT Stories from Audit Logs using Subgraph \\ Anomaly Detection and LLMs]{OCR-APT: Reconstructing APT Stories from Audit Logs using Subgraph Anomaly Detection and LLMs}

\author{Ahmed Aly}
\affiliation{%
  \institution{Concordia University}
  \city{Montreal}
  \state{Quebec}
  \country{Canada}
  }
\email{ahmed.aly.20211@mail.concordia.ca}

\author{Essam Mansour}
\affiliation{%
  \institution{Concordia University}
  \city{Montreal}
  \state{Quebec}
  \country{Canada}
}
\email{essam.mansour@concordia.ca}

\author{Amr Youssef}
\affiliation{%
  \institution{Concordia University}
  \city{Montreal}
  \state{Quebec}
  \country{Canada}
  }
\email{youssef@ciise.concordia.ca}




\begin{abstract}
Advanced Persistent Threats (APTs) are stealthy cyberattacks that often evade detection in system-level audit logs. Provenance graphs model these logs as connected entities and events, revealing relationships that are missed by linear log representations. Existing systems apply anomaly detection to these graphs but often suffer from high false positive rates and coarse-grained alerts. Their reliance on node attributes like file paths or IPs leads to spurious correlations, reducing detection robustness and reliability.
To fully understand an attack’s progression and impact, security analysts need systems that can generate accurate, human-like narratives of the entire attack. 
To address these challenges, we introduce {\sysName}, a system for APT detection and reconstruction of human-like attack stories. {\sysName} uses Graph Neural Networks (GNNs) for subgraph anomaly detection, learning behavior patterns around nodes rather than fragile attributes such as file paths or IPs. This approach leads to a more robust anomaly detection. It then iterates over detected subgraphs using Large Language Models (LLMs) to reconstruct multi-stage attack stories. Each stage is validated before proceeding, reducing hallucinations and ensuring an interpretable final report. Our evaluations on the DARPA TC3, OpTC, and NODLINK datasets show that {\sysName} outperforms state-of-the-art systems in both detection accuracy and alert interpretability. Moreover, {\sysName} reconstructs human-like reports that comprehensively capture the attack story. 
\end{abstract}

\ccsdesc[500]{Security and privacy~Intrusion detection systems}

\keywords{Anomaly Detection, APT Attack Investigation, LLMs, GNNs}

\maketitle

\renewcommand{\thefootnote}{}
\footnotetext{© 2025 Copyright held by the authors. This is the authors’ extended version of the paper accepted for publication at the ACM SIGSAC Conference on Computer and Communications Security (CCS 2025). The final published version is available at \url{https://doi.org/10.1145/3719027.3765219}.}
\renewcommand{\thefootnote}{\arabic{footnote}} 

\section{Introduction}
\label{introduction}

Advanced Persistent Threats (APTs) are among the most insidious forms of cyberattacks. Characterized by stealth, persistence, and adaptability, APTs often evade traditional security mechanisms by exploiting zero-day vulnerabilities and maintaining long-term access through low-profile tactics~\cite{alshamrani2019survey}. As a result, detecting and reconstructing these attacks from system-level audit logs remains a significant challenge for security analysts. Provenance graphs—structured representations of system logs that encode interactions between processes, files, and network entities—offer a promising way to visualize the causal relationships between system activities~\cite{lv2022review, zipperle2022provenance}. However, the complexity and size of these graphs make human analysis infeasible without intelligent automation.

Security analysts not only require systems that can detect suspicious behaviors but also demand tools that support forensic investigation by reconstructing the complete attack story. Such reconstructions must provide interpretable, human-like reports that map to APT attack stages. Existing systems typically fall short: they generate fragmented outputs or overly technical graphs that are difficult to parse and interpret. This gap motivates the need for more robust and intelligible solutions that go beyond isolated alerts and provide comprehensive insights into how an attack unfolded.

\paragraph{Limitations of Existing Detection Methods:}
Most prior efforts in APT detection using provenance data fall into two broad categories: heuristic-based and anomaly-based approaches~\cite{inam2022sok}. 
Heuristic methods rely on signatures or rules derived from known attacks~\cite{milajerdi2019holmes, hassan2020tactical}, but fail to detect novel threats.
Anomaly-based methods, in contrast, identify deviations from expected behavior and thus hold greater promise for detecting zero-day attacks~\cite{inam2022sok, wang2022threatrace}.
However, they frequently suffer from high false positive rates~\cite{unicorn, rehman2024flash}, producing voluminous alerts that burden security teams with triage tasks.
Furthermore, anomaly-based systems often operate at the node level~\cite{yang2023prographer, jia2024magic, wang2022threatrace} or over the entire graph~\cite{manzoor2016fast, kapoor2021prov, unicorn}, which creates practical limitations. Node-level alarms lack contextual information, making it difficult to interpret isolated anomalies. Graph-level alarms, on the other hand, are too coarse, obscuring the specific sequences and entities involved in an attack. Recent efforts~\cite{rehman2024flash, li2023nodlink} have shifted toward subgraph-based anomaly detection to strike a balance between granularity and interpretability. These systems identify small, connected sets of anomalous nodes to support better investigation. However, they often rely heavily on fragile node attributes like file paths or IPs, which are easy to obfuscate or manipulate, thereby reducing detection robustness~\cite{arp2022and, mukherjee2023evading}.

\paragraph{Challenges in Attack Story Reconstruction:}
Beyond detection, reconstructing a coherent and human-understandable attack story remains a major unsolved challenge. Many existing systems~\cite{atlas, fang2022back, xu2022depcomm} assume prior knowledge in the form of Points-of-Interest (POIs), such as manually flagged alerts or indicators. This reliance hinders comprehensive log analysis. Moreover, the outputs of these systems often consist of dense graphs or low-level event sequences that lack narrative clarity. Without proper summarization or contextual linking of events to known attack stages—such as those in the MITRE ATT\&CK or APT kill-chain frameworks—these systems fail to serve the needs of analysts conducting forensic investigations.

\paragraph{Our Approach:}
To address the above challenges, we propose {\sysName}\footnote{\sysName: \textbf{O}ne-\textbf{C}lass \textbf{R}elational graph convolutional networks for \textbf{APT} anomaly detection}, a novel system that performs end-to-end reconstruction of APT stories from audit logs. 
{\sysName} consists of two key components: a GNN-based {\detector}, and an LLM-based {\investigator} that generates interpretable attack stories. 
The {\detector} leverages a custom graph learning model, {\gnnModel}, which integrates relational graph convolutional networks (RGCNs) with one-class SVMs. This design captures behavioral patterns over structural relationships, allowing the system to detect anomalies based on context rather than brittle attributes. 
By training a separate model per node type, {\gnnModel} identifies abnormal interactions with higher precision.

Detected anomalous nodes are then grouped into subgraphs based on topological and behavioral coherence. Each subgraph is scored for abnormality and filtered to retain those with high investigative value. These subgraphs serve as the basis for the second component: the {\investigator}. This module applies a Retrieval-Augmented Generation (RAG) approach to serialize the subgraphs and pass them to a Large Language Model. By modularizing the reconstruction process into validated subtasks, {\sysName} mitigates common issues like LLM hallucination~\cite{perkovic2024hallucinations}. The output is a structured, stage-wise attack report that identifies Indicators of Compromise (IOCs) and maps events to the APT kill-chain~\cite{milajerdi2019holmes}.

\sloppy
\paragraph{Impact and Evaluation.} 
We evaluate {\sysName} on three provenance graph datasets: DARPA TC3~\cite{darpa-tc3}, OpTC~\cite{darpa-optc}, and NODLINK~\cite{li2023nodlink}. DARPA TC3 and OpTC are widely recognized benchmarks that reflect realistic, enterprise-scale APT scenarios~\cite{zipperle2022provenance}. The NODLINK dataset provides a controlled simulation environment for multi-stage APTs, enabling direct comparison with state-of-the-art subgraph-based anomaly detection systems.
Experimental results demonstrate that {\sysName} consistently outperforms existing systems in both anomaly detection accuracy and the interpretability of generated alerts. Specifically, {\sysName} achieves an average F1-score of 0.96, outperforming both NODLINK (0.248) and FLASH (0.945), thereby advancing the state of subgraph-based detection. Its performance is also comparable to or exceeds that of node-level and time window-based anomaly detection methods.

Beyond quantitative gains, {\sysName} advances usability by producing concise, human-readable attack reports that reconstruct a majority of the APT kill-chain stages. This capability bridges the gap between low-level system telemetry and high-level analyst reasoning. By combining graph learning with natural language generation, {\sysName} delivers not only accurate detections but also actionable insights—streamlining alert triage, reducing investigation time, and pushing the state of the art in APT detection and analysis.
Our contributions can be summarized as follows:
\begin{itemize}
\item We propose a GNN-based anomaly detection model combined with a one-class classification to accurately identify anomalous nodes and APT-related subgraphs in provenance graphs. \shorten

\item We introduce an LLM-driven investigation method that reconstructs attack stories from audit logs and generates concise, human-like reports.

\item We integrate these components into {\sysName}\footnote{Repository for {\sysName}: \url{https://github.com/CoDS-GCS/OCR-APT}}, a complete APT detection and investigation system that identifies anomalies, ranks alerts by severity, and produces interpretable reports to support efficient analyst workflows.

\item We conduct extensive evaluations on DARPA TC3, OpTC, and NODLINK datasets. {\sysName} outperforms state-of-the-art anomaly detection systems and successfully reconstructs multi-stage, human-like APT reports.
\end{itemize}

The remainder of the paper is organized as follows: Section~\ref{background} reviews background and limitations of existing systems; Section~\ref{threat-model} defines the threat model. Section~\ref{system-architecture} introduces our system, with Sections~\ref{gnn-detector} and~\ref{llm-investigator} detailing the GNN-based detector and LLM-based {\investigator}, respectively. Evaluation results are in Section~\ref{evaluation}, related work in Section~\ref{related-work}, and conclusions in Section~\ref{conclusion}.

\section{Background}
\label{background}

Provenance graphs (PGs) are directed, heterogeneous graphs that model audit logs to support causal analysis~\cite{lv2022review, zipperle2022provenance}. They provide a comprehensive view of system activities and information flow, making them effective for uncovering attack traces~\cite{inam2022sok}.
A PG consists of diverse node types—such as processes, files, and network flows—linked by edges representing actions like read, write, and execute. The exact schema depends on the underlying operating system; our approach leverages all available node and edge types per host.
For example, Appendix~\ref{appendix-datasets} outlines the schema used for FreeBSD-based hosts (CADETS).
PGs also include event timestamps, crucial for detecting APTs and reconstructing attack timelines~\cite{rehman2024flash}, as well as contextual node attributes like command-line arguments, file paths, and IP addresses. These features enrich the analysis of system behavior.

\subsection{Limitations of Anomaly Detection Systems}
\label{existing-systems}

Anomaly-based detection systems learn patterns of normal system behavior and flag deviations as potential threats. In provenance graph analysis, early approaches often detect anomalies at the granularity of entire graphs using clustering~\cite{manzoor2016fast, unicorn} or graph classification techniques~\cite{kapoor2021prov, huang2022one}. However, these methods struggle with interpretability: the alarms often span PGs with millions of nodes~\cite{inam2022sok}, despite only a small subset being relevant to the attack~\cite{rehman2024flash, li2023nodlink}. This makes the investigation akin to “searching for a needle in a haystack.” Additionally, such coarse-grained approaches risk missing fine-grained anomalies, leading to false negatives~\cite{wang2022threatrace}.

To improve granularity, subsequent systems focus on paths~\cite{provdetector, liu2019log2vec}. While more precise, these techniques often lose broader attack context~\cite{hassan2020tactical}. More recent advancements target even finer units—individual nodes~\cite{wang2022threatrace, jia2024magic, yang2023prographer}, edges~\cite{zengy2022shadewatcher}, time windows~\cite{cheng2024kairos}, or subgraphs~\cite{li2023nodlink, rehman2024flash}. Although these methods enhance interpretability, node- and edge-level systems can overwhelm analysts with numerous isolated alerts lacking contextual history.

Subgraph-based systems like NODLINK~\cite{li2023nodlink} attempt to offer better context by constructing coherent attack graphs using Steiner Trees~\cite{hwang1992steiner, imase1991dynamic}. However, NODLINK's reliance on sentence embeddings~\cite{bojanowski2017enriching} limits precision, often resulting in false positives. 
Moreover, NODLINK relies on node attributes features that may introduce spurious correlations—a common issue in cybersecurity ML, where models learn artifacts (e.g., specific IP ranges) instead of generic attack patterns~\cite{arp2022and}.
FLASH~\cite{rehman2024flash} and KAIROS~\cite{cheng2024kairos} follow similar strategies, using GNNs with node attributes such as process names, command-line arguments, file paths, and IP addresses to inform embeddings. While these semantic features improve detection accuracy, they are also vulnerable to adversarial manipulation, as attackers can change surface-level attributes without altering attack behavior~\cite{mukherjee2023evading}.
To counteract this vulnerability, {\sysName} takes a novel approach by avoiding reliance on node attributes. Instead, it uses structural and behavioral features to strengthen robustness against evasion tactics. This strategy ensures consistent anomaly detection, making {\sysName} more accurate and reliable than existing subgraph-level systems, which still struggle with false positives and adversarial manipulation.

\paragraph{Methodological Limitations.} 
Current anomaly detection methods commonly rely on two paradigms: autoencoders~\cite{jia2024magic, yang2023prographer, li2023nodlink, cheng2024kairos} and node-type classification~\cite{wang2022threatrace, rehman2024flash}. Autoencoders are memory-intensive due to the need to reconstruct large adjacency matrices~\cite{liu2022bond}. 
In node-type classification, a node is flagged as anomalous if its predicted type (e.g., process, file, network flow) differs from the expected one. However, this assumption does not always hold, as each node type exhibits distinct behavioral patterns. For example, process nodes perform distinct actions that reveal their type, so a malicious process may still be correctly classified and evade detection.
{\sysName} mitigates this issue by avoiding type-based classification. Instead, it directly classifies nodes as normal or anomalous based on their behavioral patterns, using a one-class SVM~\cite{bounsiar2014one}. This one-class classification approach identifies outliers without relying on labeled attack data, enabling the detection of previously unseen attack behaviors.

Moreover, prior works~\cite{wang2022threatrace, rehman2024flash} employ GNN models originally designed for homogeneous graphs, such as GraphSAGE~\cite{hamilton2017inductive}, which do not consider edge types (i.e., node actions) during embedding computation. As a result, they overlook critical structural context. While GNNs have been actively explored for anomaly detection, the heterogeneous nature of PGs remains underexplored~\cite{kim2022graph}.
{\sysName} addresses this gap by using RGCNs~\cite{rgcn} to embed nodes while preserving the heterogeneous structure of PGs. Unlike previous methods, {\sysName} incorporates node actions directly into node embeddings, which allows it to account for complex relationships between nodes in attack scenarios. 
Section~\ref{components-analysis} compares {\sysName} with other GNN-based baselines, demonstrating its superior performance in terms of both precision and recall.

\paragraph{Efficiency Considerations.} 
Scalability is critical for handling large-scale enterprise provenance data. While prior systems propose graph reduction and subgraph extraction techniques~\cite{aly2024megr, hossain2017sleuth, altinisik2023provg}, {\sysName} introduces a memory-efficient approach tailored specifically to anomaly detection. Instead of extracting graphs from known IOCs, {\sysName} constructs causally relevant subgraphs around detected anomalous nodes using three efficient graph queries. This method avoids the need to load the entire PG into memory, which is essential for supporting deployment in resource-constrained environments and ensuring scalability.

\subsection{Limitations of Attack Investigation Systems}

Attack investigation systems support post-alert analysis by helping security analysts validate threats and understand the attacker’s actions~\cite{inam2022sok}. Key challenges include triaging high-priority alerts~\cite{hassan2019nodoze, hassan2020we}, clustering related alerts~\cite{van2022deepcase, zeng2021watson}, and reconstructing comprehensive attack stories from low-level logs~\cite{atlas, fang2022back, xu2022depcomm, li2023nodlink, cheng2024kairos}.

Many reconstruction systems rely on pre-identified POIs~\cite{fang2022back, xu2022depcomm, van2022deepcase} or known attack entities~\cite{atlas} as seeds for investigation. This dependence limits generalization: if the initial POI is inaccurate, the derived attack story may be misleading. For instance, ATLAS~\cite{atlas} trains an LSTM-based model on simulated attack sequences. If the starting entity is misclassified, the entire reconstruction may be compromised. Such systems are often unable to detect novel threats outside the scope of their training data.
Rule-based systems~\cite{milajerdi2019holmes, hossain2017sleuth} suffer similar drawbacks, as predefined patterns can only capture known attacks. They are ineffective against polymorphic APTs that exhibit diverse behaviors~\cite{inam2022sok, unicorn, wang2022threatrace}.

\paragraph{Narrative Complexity.} 
Most existing tools generate either attack graphs~\cite{xu2022depcomm, li2023nodlink, steinerlog, pei2016hercule, hossain2017sleuth, milajerdi2019holmes} or attack sequences~\cite{atlas, fang2022back}. While informative, these representations are often complex and hard to interpret, requiring significant analyst effort to extract key insights. Graphs, in particular, pose challenges for visualization and manual inspection~\cite{cheng2024kairos}.
Some approaches focus on alert clustering~\cite{van2022deepcase, liu2022rapid, zeng2021watson}, grouping similar alerts to reduce manual workload. However, they typically lack narrative coherence and contextual depth, which are essential for understanding multi-stage APTs.
In contrast, {\sysName} introduces a novel LLM-based module that generates high-quality, human-like attack reports, offering coherent summaries of attack behavior. These reports not only reduce analyst burden but also improve the effectiveness and speed of APT investigations by providing a clear, interpretable story that can easily guide further analysis and response.

\section{Threat Model}
\label{threat-model}
This study focuses on detecting APTs characterized by a “low and slow” attack approach~\cite{unicorn}. While our threat model acknowledges that attackers may use sophisticated zero-day exploits to compromise the system, we assume they must leave distinguishable traces in the system logs.
The proposed system requires system logs that are free from attack traces for training.
Consistent with previous work, we consider audit logs and kernel-space auditing frameworks as part of our trusted computing base~\cite{atlas, milajerdi2019holmes, zengy2022shadewatcher,aly2024megr,jia2024magic}. 
Attacks involving data poisoning, hardware-level attacks, and side-channel attacks are beyond the scope of this study.

\begin{figure*}[t]
  \centering  
  \includegraphics[width = \linewidth]{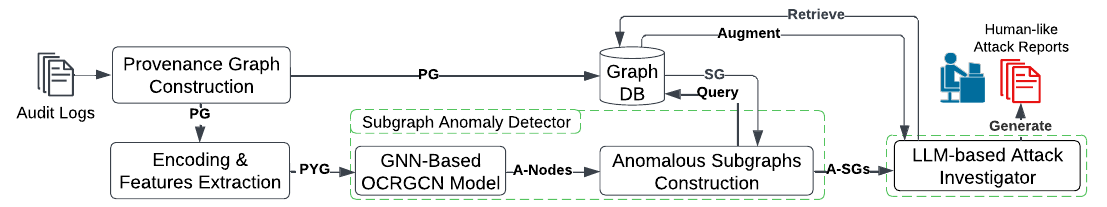}
  \caption{Overall architecture of {\sysName}. This includes constructing provenance graphs (PG), extracting features and encoding the graph into a PyTorch Geometric data object (PYG), detecting anomalous nodes (A-Nodes) with our GNN-based model (OCRGCN), identifying anomalous subgraphs (A-SGs), and generating a human-like APT attack report using LLMs.}
  \label{fig:ocrapt-system-pipeline}
\end{figure*}

\section{Proposed System Architecture}
\label{system-architecture}
This section presents the architecture of our {\sysName} system, shown in Figure~\ref{fig:ocrapt-system-pipeline}. Each component of {\sysName} is designed to address key limitations identified in existing anomaly detection and attack investigation systems (Section~\ref{existing-systems}), and together they achieve our research objectives: fine-grained anomaly detection, resistance to adversarial evasion, scalability to enterprise-scale data, and automated, interpretable threat investigation.

To address the scalability and memory-efficiency limitations of existing systems, {\sysName} represents audit logs as an RDF-based\footnote{RDF: Resource Description Framework graph model~\cite{world2014rdf}} provenance graph (PG) and loads them incrementally into an RDF graph database using a disk-backed ingestion strategy~\cite{aly2024megr}. The RDF model encodes system activity as triples of the form \RDFTYPE{subject}{predicate}{object}, where \texttt{predicate} denotes an event, and \texttt{subject} and \texttt{object} are system entities. This representation enables scalable graph construction and efficient query-based subgraph extraction, overcoming the limitations of systems that require the entire graph to be held in memory.

To address the interpretability challenges of coarse-grained anomaly detection, {\sysName} extracts behavior-based features for each node, including actions, effects, and timing statistics. Normalization ensures consistency across entities, supporting generalized behavior learning. The graph is then encoded for GNN-based modeling.
The encoded graph is passed to the {\detector}, which overcomes limitations of prior methods, such as reliance on attribute embeddings or autoencoders, by combining node-level anomaly detection with context-aware subgraph analysis. Using {\gnnModel}, a one-class GNN trained on benign data, it identifies anomalous nodes based on structural and behavioral features without labeled attacks. To improve interpretability and reduce alert fatigue, {\sysName} constructs causally coherent subgraphs around detected anomalies using efficient SPARQL queries. Each subgraph is scored, and those above a threshold are flagged, providing precise and context-rich alerts without static rules.

To support effective post-alert investigation and narrative reconstruction, anomalous subgraphs are passed to the {\investigator} module, which uses LLMs to generate concise, human-readable attack reports. It serializes each subgraph into a timestamp-ordered description, summarizes it via LLM prompts to extract IOCs, key actions, and APT kill chain stages, and then composes a complete attack narrative. This report is further enriched through a retrieval-augmented generation (RAG) pipeline that queries the graph for additional context.
By integrating precise anomaly detection with automated investigation, {\sysName} reduces false positives, enhances interpretability, and scales to large, heterogeneous provenance graphs—effectively addressing the core challenges outlined in our research objectives.


\section{GNN-based Subgraph Anomaly Detection}
\label{gnn-detector}

This section introduces the {\detector}, outlining the {\gnnModel} architecture and subgraph construction algorithm. 

\subsection{The GNN-based model}
\begin{figure*}[t]
  \centering
  \includegraphics[width = \linewidth]{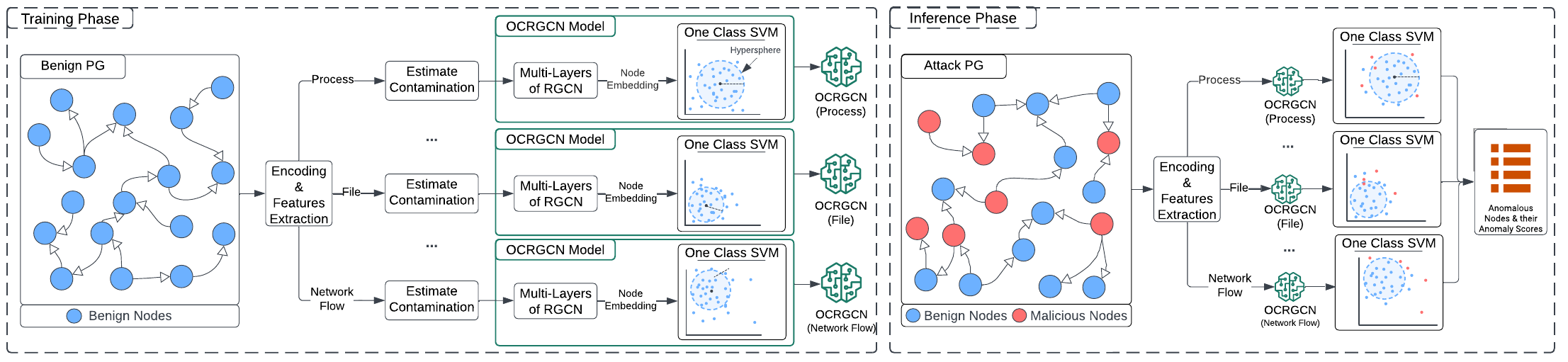}
  \caption{Architecture of the {\gnnModel} model. The training phase (left) involves encoding benign provenance graphs, estimating contamination factors, learning node embeddings via RGCN layers, and learning the normal hypersphere with a one-class SVM. The inference phase (right) uses the trained models to compute anomaly scores and detect anomalous nodes.}  
  \label{fig:OCRGCN_model}
\end{figure*}

The GNN-based model acts as the core component of {\sysName}'s subgraph anomaly detection pipeline. Figure~\ref{fig:OCRGCN_model} illustrates the model's architecture, highlighting its training and inference phases.

\subsubsection{The model architecture} 
Our {\gnnModel} uses an RGCN-based architecture~\cite{rgcn} to capture both graph structure and node behavior by incorporating edge types during the embedding aggregation process. 
To prevent spurious correlations~\cite{arp2022and}, {\gnnModel} avoids using node attributes such as IP addresses or file paths, which can cause the model to memorize specific malicious instances rather than learn generalizable attack patterns. 
While excluded from the model's input, these attributes are retained for the investigation phase, where they assist analysts in interpreting and verifying alerts.
Each layer of the model aggregates information from a node’s one-hop neighborhood. Nodes exchange messages with their neighbors, embedding information about node types, actions, and initial extracted features, and then update their embeddings based on the aggregated data. After multiple RGCN layers, the model produces a final embedding vector for each node in the provenance graph.
These embeddings are then passed to a one-class SVM, which learns a hypersphere that encloses the majority of normal node embeddings. An anomaly score is computed based on the distance of each node’s embedding from the center of this hypersphere. Nodes whose scores exceed the hypersphere’s radius are flagged as anomalous.

\subsubsection{Training Phase} 
The training begins by extracting behavior-based features from benign provenance graphs, including action frequencies and idle period statistics. 
Action frequencies reflect behavioral tendencies by computing the proportion of each action type relative to the total number of actions per node. Unlike prior work that uses raw action counts~\cite{wang2022threatrace}, we apply L2 normalization to reduce bias from high-activity benign nodes and allow the model to focus on behavioral patterns (i.e., scaling the action frequency vector so that the sum of squared values equals one).
For example, while a frequently used browser may establish many connections, its overall behavior is normal when considering the ratio of connections to other actions like sends, receives, reads, and writes. Conversely, a process with an unusually high rate of sending or execution may indicate suspicious behavior.
Idle period statistics are derived from event timestamps and include minimum, maximum, and average durations between actions. These statistics are normalized to a 0–1 range using a min-max scaler, based on the dataset’s minimum and maximum values. APT-related nodes tend to remain idle longer than benign nodes, making idle period statistics a key indicator.
These two features capture key aspects of node behavior and assist in detecting malicious patterns. Importantly, {\sysName} does not rely solely on these raw features; its GNN aggregates them within the graph structure, enabling node representations to reflect both behavioral patterns and structural context. We conducted additional experiments to evaluate alternative temporal features, but ultimately excluded them due to poor generalization across hosts; further details are provided in Appendix~\ref{feature-selection-appx}.

After feature extraction, the system splits nodes by type and trains a specific {\gnnModel} for each type to improve detection accuracy, as normal behavior varies across node types. 
Each {\gnnModel} is trained on a single node type but aggregates messages from all neighboring types, preserving cross-type semantics. Following this, each model learns a hypersphere specific to a given node type, enabling anomaly detection tailored to the normal behavior of that type without losing cross-type interaction information.
Distinct models also enable precise estimation of the contamination factor, which represents the expected proportion of anomalies. This factor is estimated as the proportion of malicious nodes in the validation set, constrained between $Min_{con}$ and $Max_{con}$.
The maximum constraint ensures the contamination factor aligns with the stealthy nature of APTs. If the validation set contains many malicious nodes, the contamination factor is set to $Max_{con}$. The minimum constraint ensures the factor is above zero, even when no malicious nodes are present. If no labeled data is available, the system uses $Min_{con}$ as the contamination factor, relying solely on trusted benign logs reflecting normal behavior.

Each {\gnnModel} model learns a hypersphere that encloses most normal nodes for a specific type and computes the anomaly score threshold based on the contamination factor. The fraction of training nodes allowed outside the hypersphere is controlled by the hyperparameter $\beta$, fixed across all node types. If too many nodes are enclosed, the hypersphere becomes too large, reducing its ability to detect anomalies. The goal is to capture the norm of benign nodes without overfitting.
During training, the RGCN model updates its weights to bring normal nodes closer together in the embedding space, while the one-class SVM adjusts the hypersphere’s center and radius to fit the normal node embeddings. Training stops early based on the validation set’s F1-score, and if no malicious nodes are present, training halts when the true negative rate declines.

\begin{figure*}[t]
  \centering
  \includegraphics[width = \linewidth]{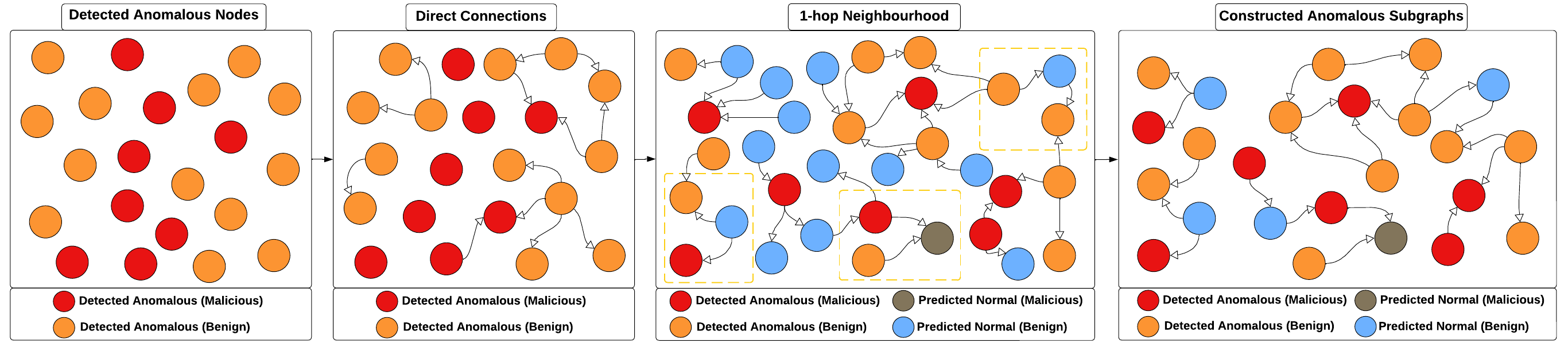}
  \caption{The stages of constructing anomalous subgraphs. {\sysName} starts from anomalous nodes, connects them by direct connections, gets all one-hop neighbor nodes, and keeps only neighbors that lead to other anomalous nodes.}
  \label{fig:construct_subgraphs}
\end{figure*}

\subsubsection{Inference Phase} During inference, the system processes provenance graphs containing both benign and malicious traces. It applies the same pre-processing steps to extract features and assigns each node to its corresponding {\gnnModel} model based on its type. The {\gnnModel} models compute anomaly scores for the test nodes, and those exceeding the pre-computed threshold are classified as anomalous. Finally, these anomalous nodes and their scores are passed to the subgraph construction module.

\begin{algorithm}[t]
	\caption{Anomalous Subgraphs Construction Algorithm}
	\label{subgraph-construction-algorithm}
	\begin{algorithmic}[1]
            \STATE \textbf{Input:} Provenance Graph Database ($DB_{PG}$), Anomalous Nodes ($A_{Nodes}$), $n_{seed}$, $max_{e}$
            \STATE \textbf{Output:} Anomalous Subgraphs ($A_{SGs}$)
            \STATE Query direct connections between $A_{Nodes}$ from $DB_{PG}$ 
            \STATE Query 1-hop neighbors of $A_{Nodes}$ from $DB_{PG}$ 
            \STATE Construct an initial subgraph ($init_{SG}$)
            \STATE Sort $A_{Nodes}$ based on their anomaly scores  
            \STATE Identify $Seeds$ as top $n_{seed}$ $A_{Nodes}$ per node type 
            \FOR {every $Seeds$}
            \STATE Traverse $init_{SG}$ for 1-hop forward and backwards
            \STATE Keep only paths that lead to unvisited $A_{Nodes}$
            \STATE Construct a subgraph ($sg$) 
            \IF {$sg_{edges}<=Max_{e}$}
            \STATE Add $sg$ to to the Anomalous Subgraphs ($A_{SGs}$) list
            \ELSE 
            \STATE Partition $sg$ into smaller subgraphs within $Max_{e}$
            \STATE Add partitioned subgraphs to $A_{SGs}$
            \ENDIF
            \ENDFOR
            \STATE Filter out identical subgraphs in $A_{SGs}$
            \FOR{every $A_{SGs}$}
            \STATE Compute the subgraph anomaly score
            \STATE Determine the subgraph abnormality level ($sg_{ab}$)
            \ENDFOR
            \STATE Filter out subgraphs with minor $sg_{ab}$ from $A_{SGs}$
	\end{algorithmic} 
\end{algorithm}

\subsection{Anomalous Subgraph Construction} 
\label{subgraph-construction-algo}
{\sysName} constructs anomalous subgraphs using Algorithm~\ref{subgraph-construction-algorithm}. The algorithm takes as input a set of anomalous nodes, a connection to the provenance graph database, and two parameters: $n_{seed}$, which specifies the number of seed nodes for each node type, and $max_e$, the maximum number of edges allowed in each subgraph. It begins by querying the graph database to retrieve direct connections between anomalous nodes and their one-hop neighbors to form an initial subgraph (lines 3-5). This is done with three SPARQL queries: one for direct edges between anomalous nodes, and two for their neighboring nodes and edges, minimizing traversal overhead.

The nodes are next ranked by anomaly scores, and the top $n_{seed}$ nodes of each type are selected as seeds (lines 6–7). For each seed, the algorithm performs a 1-hop bidirectional traversal (line 9), connecting anomalous nodes through intermediate normal ones to preserve their context. It then retains only the paths that lead to anomalous nodes and constructs a candidate subgraph (lines 10–11), limiting the inclusion of benign nodes.
Figure~\ref{fig:construct_subgraphs} illustrates this process: it begins with individual anomalous nodes, links them via direct connections when possible, expands each by one hop to capture surrounding context, and then prunes paths that do not lead to additional anomalies. The figure shows how anomalous nodes (in red and orange) are connected through normal nodes (in blue), with misclassified nodes (in brown) also identified.

Once a candidate subgraph is formed, it is either added directly to the set of anomalous subgraphs if it stays within the edge limit $max_e$, or partitioned into smaller subgraphs (lines 12–17). The Louvain community detection algorithm~\cite{blondel2008fast} is used for partitioning. This ensures manageable subgraph sizes for analysis and helps the LLM-based attack investigation maintain narrative coherence. 
Partitioning does not affect investigation quality, as the LLM-based investigator summarizes each partition individually and merges them into a comprehensive attack report.
Some cross-partition edges may be omitted, but the overall attack scenario remains intact.
After processing all seed nodes, duplicate subgraphs are removed (line 19) and an anomaly score is computed for each subgraph (line 21). This score is the sum of the scores of its anomalous nodes. The subgraphs are then mapped to abnormality levels using a logarithmic scale (line 22), and those with low abnormality are filtered out (line 24), reducing false positives. The final set of anomalous subgraphs can be adjusted based on the desired abnormality threshold for further investigation.

\begin{figure*}[t]
  \centering
  \includegraphics[width = \linewidth]{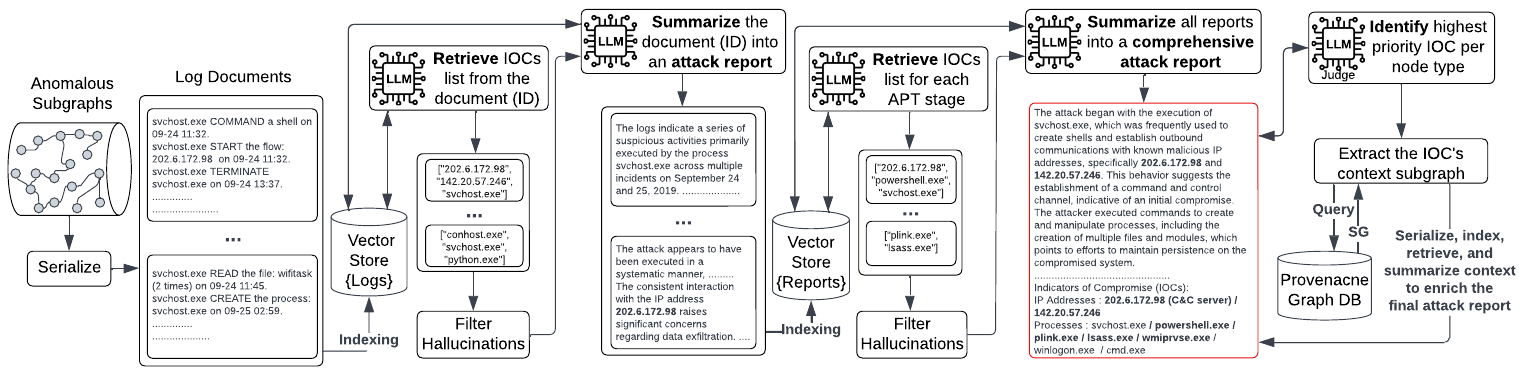}
  \caption{Architecture of the LLM-based {\investigator}. The system serializes anomalous subgraphs into log documents, indexes them in a vector store, and uses an LLM to generate attack reports. It identifies key IOCs, enriches reports with context subgraphs, and produces a comprehensive report for analysts. The visualized reports are simplified versions of the recovered report from host 501 in the DARPA OpTC dataset.}
  \label{fig:attack_investigator}
\end{figure*}


\section{LLM-Based Attack Investigation}
\label{llm-investigator}

\subsection{The Limits of LLMs in Attack Investigation}

To enable LLM-based attack investigation, we examined how LLMs perform when tasked with reconstructing attack stories from system audit logs. This reconstruction process demands high-level reasoning, contextual understanding, and the interpretation of subtle event patterns. We found that LLMs struggle to generate high-quality of human-like reports when asked to perform this task in a single step. Our initial approach used a monolithic prompt to generate full reports directly from anomalous subgraphs, but this often led to hallucinated content, overlooked APT stages, and missing IOCs. These limitations persisted even when evaluating on benchmark datasets that may have been seen during pretraining. This shows the inherent difficulty of complex and analyst-level investigative tasks when attempted all at once.

To address these challenges, we first incorporated Chain-of-Thought (CoT) prompting~\cite{wei2022chain}, embedding explicit reasoning steps to help the model logically interpret each subgraph. This improved the coherence of the report and increased the coverage of the APT stage, but hallucinations and factual errors remained. We then designed a multi-stage prompting pipeline that decomposes the investigation into smaller, well-defined subtasks, such as IOC extraction, APT stage mapping, and context summarization. 
Each stage employs a focused CoT-based prompt, enhanced by in-context learning with domain-specific instructions and CTI concepts. This enables the LLM to reason more effectively within a constrained scope.

Building on these insights, we designed a complete LLM-driven attack investigation mechanism, called {\investigator}. It implements our multi-stage prompting strategy within a Retrieval-Augmented Generation (RAG) framework. 
Each stage uses tailored CoT-based prompts and is connected by an automatic validation mechanism that ensures consistency and preserves report integrity.
This design not only improves investigative performance by capturing more attack stages, but also fully mitigates hallucinations observed in the earlier approach. 
This modular pipeline, with explicit reasoning and automatic validation, improves the overall quality and completeness of the generated attack reports. Hence, our approach enables reliable and context-aware attack report generation that accelerates attack investigations.

\subsection{Our Attack Investigator Mechanism}

We have designed the {\investigator} to reconstruct attacks via a RAG-based pipeline consisting of six stages, as shown in Figure~\ref{fig:attack_investigator}. The pipeline transforms detected anomalous subgraphs into human-like attack reports, which capture the main stages of the attack story. In the first stage, anomalous subgraphs are serialized into event log documents and indexed in a vector store. These serialized logs provide a structured representation of the subgraphs for further processing.
Stage two involves using an LLM to extract IOCs from each serialized subgraph stored in the vector store. 
To prevent hallucinations, the system validates the extracted IOCs by checking whether they appear in the corresponding anomalous subgraphs. This ensures that only verified IOCs are retained for subsequent stages.
In stage three, the LLM generates an attack report for each subgraph based on the validated IOCs. These reports are indexed into a separate vector store, making them easily retrievable. The final comprehensive attack report is reconstructed in stages four and five. In stage four, the LLM extracts a list of IOCs for each APT stage from all generated attack reports. These individual reports are merged into a comprehensive attack report, as shown in the red box in Figure~\ref{fig:attack_investigator}.

The final stage enriches the comprehensive report by iterating over the most critical IOCs. The system employs a mechanism called llm-as-a-judge~\cite{zheng2024judging} to identify the most significant IOCs. The system then queries the provenance graph database to retrieve the contextual information for each identified IOC, which is in the form of connected anomalous subgraphs. Each subgraph provides crucial attack context, which is integrated into the comprehensive report. This process enhances the detection of additional APT stages.


\begin{algorithm}[t]
    \caption{Attack Reports Generation Algorithm}
    \label{report-generation-algorithm}
    \begin{algorithmic}[1]
        \STATE \textbf{Input:} Provenance Graph Database ($DB_{\text{PG}}$), Anomalous Subgraphs ($ASG_{s}$)
        \STATE \textbf{Output:} Attack Reports ($R_{atk}$), Comprehensive Attack Report ($R_{\text{comp}}$)  
        \STATE Serialize $ASG_{s}$ into log documents ($ASG_{\text{docs}}$)
        \STATE Index $ASG_{\text{docs}}$ in a vector store for logs ($VSt_{\text{logs}}$)
        \STATE Initialize LLM chat engine ($LLM_{\text{chat}}$) with instructions
        \FOR {each $ASG_{\text{doc}}$ in $ASG_{\text{docs}}$}
            \STATE Extract $IOC_{\text{lst}}$ using $LLM_{\text{chat}}$ from $VSt_{\text{logs}}$
            \STATE Filter hallucinations in $IOC_{\text{lst}}$
            \STATE Summarize $ASG_{\text{doc}}$ using $LLM_{\text{chat}}$, append into $R_{\text{atk}}$
            \STATE Reset $LLM_{\text{chat}}$ memory
        \ENDFOR
        \STATE Index $R_{\text{atk}}$ in a vector store for reports ($VSt_{R}$)
        \STATE Extract $IOC_{\text{lst}}$ per APT stage from $R_{\text{atk}}$ using $LLM_{\text{chat}}$
        \STATE Filter hallucinations in $IOC_{\text{lst}}$
        \STATE Summarize $R_{\text{atk}}$ into $R_{\text{comp}}$ using $LLM_{\text{chat}}$
        \STATE Initialize LLM as a judge ($LLM_{\text{judg}}$) with instructions
        \FOR {each $node_{\text{type}}$ in [`IP', `PROCESS', `FILE']}
            \STATE Prompt $LLM_{\text{judg}}$ to select most critical $IOC$ in $R_{\text{comp}}$
            \STATE Query $DB_{\text{PG}}$ to retrieve $IOC$ context
            \STATE Serialize $IOC$ context and index it in $VSt_{\text{logs}}$
            \STATE Extract $IOC_{\text{lst}}$ per APT stage using $LLM_{\text{chat}}$
            \STATE Filter hallucinations in $IOC_{\text{lst}}$
            \STATE Summarize context using $LLM_{\text{chat}}$, append into $R_{\text{atk}}$
            \STATE Enrich $R_{\text{comp}}$ with the report using $LLM_{\text{chat}}$
        \ENDFOR
    \end{algorithmic}
\end{algorithm}

\subsection{Attack Report Generation}

The {\investigator} generates attack reports using Algorithm~\ref{report-generation-algorithm}, which takes anomalous subgraphs as input and produces a comprehensive narrative that reconstructs the attack story at the subgraph level. Each detected anomalous subgraph represents a fragment of the broader attack scenario.
The algorithm begins by serializing each subgraph into a chronological sequence of events (line 3). This process converts the subgraph's edges into natural language sentences that encode subject and object attributes, the action performed, and the associated timestamps. During serialization, a reduction phase condenses duplicate actions occurring within one-second intervals into a more compact representation. For instance, if a process repeatedly reads the same file, the output records the action once, noting the number of repetitions (e.g., “read X times”). Timestamps are simplified from microseconds to seconds to reduce token overhead and streamline the LLM input.

The serialized subgraph (log document) is segmented into sentence chunks. Embeddings for each chunk are computed and indexed into a vector store (line 4). This enables the LLM to efficiently retrieve relevant context from log documents. 
We use the `text-embedding-3-large' model~\cite{openai2024textembedding3} for indexing, due to its superior performance~\cite{cheng2024ctinexus}.
The algorithm then configures the LLM with domain-specific instructions for attack investigation (line 5). These instructions guide the model’s reasoning by narrowing its focus to concepts from operating system security, Cyber Threat Intelligence (CTI), and APT kill-chain stages. They also define key terms, such as IOCs and APT stages, emphasize the importance of avoiding hallucinations, and require the model to produce high-quality, human-like narratives. The task is framed as summarizing detection alerts into concise reports that include: (1) a summary of attack behavior, (2) a breakdown of APT stages, (3) identified IOCs with context, and (4) a minute-by-minute action log. The full set of instructions and prompts is provided in Appendix~\ref{llm-prompts}.

To enhance performance, the reconstruction process is modularized into subtasks, enabling the LLM to focus on one task at a time. First, the model extracts IOCs from the serialized document ($ASG_{\text{doc}}$) using a dedicated prompt ($p_{\text{ioc}}$) as shown in Equation~\ref{prompt:extract-ioc}:

\begin{equation}
\label{prompt:extract-ioc}
\{IOC_i\} = f(ASG_{\text{doc}}, p_{\text{ioc}}) 
\end{equation}

This modular design allows the system to validate extracted IOCs and filter out hallucinations. Specifically, any IOCs not found in the source document are excluded (line 8). The validated IOCs ($\{IOC_i\}'$) are then used to guide the LLM in generating an attack report ($R_{\text{atk}}$) using prompt $p_{\text{sum}}$ (line 9), as shown in Equation~\ref{prompt:summarize-report}. After generating each report, the system resets the LLM’s memory to avoid cross-document contamination (line 10).

\begin{equation}
\label{prompt:summarize-report}
R_{\text{atk}} = f(ASG_{\text{doc}}, \{IOC_i\}', p_{\text{sum}}) 
\end{equation}

Once all subgraphs have been processed, the reports ($R_{\text{atk}}$) are indexed, and the LLM is prompted to extract IOCs per APT stage ($stg$), supporting the creation of a unified attack report ($R_{\text{comp}}$) (lines 12–15):

\begin{equation}
\label{prompt:extract-ioc-per-stage}
\{IOC_{\text{stg}}^{(i)}\} = f(\{R_{\text{atk}}^{(i)}\}, stg, p_{\text{ioc,stg}})
\end{equation}

\begin{equation}
\label{prompt:summarize-comp}
R_{\text{comp}} = f(\{R_{\text{atk}}^{(i)}\}, \{IOC_{\text{stg}}^{(i)}\}', p_{\text{comp}})
\end{equation}

To further enrich $R_{\text{comp}}$, the system applies a RAG-based process. It begins by initializing an LLM “judge” guided by expert-level instructions (line 16). This step enables the fully automated pipeline to evaluate the generated content. The judge LLM selects the most critical IOC per node type using $p_{\text{judg}}$ (lines 17–18):

\begin{equation}
\label{prompt:select-ioc-judge}
IOC = f(R_{\text{comp}}, p_{\text{judg}})
\end{equation}

The system then queries the provenance graph to extract subgraphs centered on the selected IOCs (line 19). Graph traversal is limited to one-hop anomalous nodes, filtering out benign context—especially relevant when attackers use “living-off-the-land” techniques by exploiting legitimate system processes. Since such nodes can produce excessive benign context, the filter helps keep the investigation focused.
Finally, the extracted context subgraphs ($IOC_{\text{ctx}}$) are indexed, and used to augment the comprehensive report using prompt $p_{\text{aug}}$ (lines 20–25):

\begin{equation}
\label{prompt:augment-comp}
R_{\text{comp}}' = f(R_{\text{comp}}, IOC_{\text{ctx}}, p_{\text{aug}})
\end{equation}

The final report, $R_{\text{comp}}'$, offers an accurate reconstruction of the attack story. Analysts can interact with the system by posing follow-up questions to the LLM, such as assessing the security context of a specific entity, evaluating the likelihood of exploitation, or differentiating between malicious and benign behaviors in the subgraphs. They may also identify additional IOCs for further investigation. Overall, {\investigator} supports analysts with a user-friendly and effective interface for in-depth incident analysis.


\section{Evaluation}
\label{evaluation}
This section presents a comprehensive evaluation of {\sysName}. We compare its detection accuracy with state-of-the-art (SOTA) anomaly detection systems, excluding rule-based systems since they target known attacks, while anomaly-based methods detect novel threats.
We also study the impact of core components on accuracy via ablation and assess their computational cost. Finally, we evaluate the quality of our LLM-based investigation by comparing generated reports to ground truth reports from simulated attacks.

\subsection{Datasets} 
We evaluated our system on three datasets: DARPA Transparent Computing Engagement 3 (TC3)~\cite{darpa-tc3}, DARPA Operationally Transparent Cyber (OpTC)~\cite{darpa-optc}, and NODLINK simulated dataset~\cite{li2023nodlink}. 
These datasets consist of audit logs collected from diverse operating systems, with a total exceeding 80 million system events.
Detailed statistics for each dataset are summarized in Table~\ref{tab:datasets-statistics}. On average, malicious nodes represent less than 0.01\% of the total nodes, which aligns with typical APT behaviors. Therefore, we used the F1-score as the primary evaluation metric, as it effectively measures performance on highly imbalanced datasets~\cite{naidu2023review}.

\begin{table}[t]
\centering
\caption{Statistics of DARPA TC3, DARPA OpTC, and Simulated NODLINK datasets} 
\label{tab:datasets-statistics}
\resizebox{\columnwidth}{!}{%
\begin{tabular}{@{}cccccccc@{}}
\toprule
Dataset &
  Host &
  \# Nodes &
  \# Edges &
  \begin{tabular}[c]{@{}c@{}}\# Nodes \\ Types\end{tabular} &
  \begin{tabular}[c]{@{}c@{}}\# Edges \\ Types\end{tabular} &
  \begin{tabular}[c]{@{}c@{}}\# Malicious \\ Nodes\end{tabular} \\ 
  \midrule
\multirow{3}{*}{\begin{tabular}[c]{@{}c@{}}DARPA \\ TC3\end{tabular}}  & CADETS & 696.37 K & 8.66 M  & 6  & 28 & 12.81 K \\ 
                            & TRACE  & 2.48 M   & 6.98 M  & 11 & 24 & 67.38 K \\  
                            & THEIA  & 642.56 K & 18.82 M & 4  & 18 & 25.32 K \\ \hline
\multirow{3}{*}{\begin{tabular}[c]{@{}c@{}}DARPA \\ OpTC\end{tabular}} & 201 & 788.24 K & 5.84 M  & 8  & 20 & 71      \\ 
                            & 501 & 1.14 M   & 8.29 M  & 8  & 20 & 418     \\  
                            & 51 & 720.40 K & 4.98 M  & 7  & 19 & 200     \\ \hline
\multirow{3}{*}{\begin{tabular}[c]{@{}c@{}}Simulated \\ NodLink\end{tabular}}   & Ubuntu & 23.04 K  & 14.04 M & 3  & 13 & 21      \\ 
                            & WS12 & 10.86 K  & 8.27 M  & 3  & 6 & 47      \\ 
                            & W10 & 62.16 K  & 7.89 M  & 3  & 6 & 191     \\ 
                            \bottomrule
\end{tabular}
}
\end{table}

\subsubsection{DARPA TC3}
The DARPA TC3 dataset, widely used as a benchmark for provenance graph intrusion detection~\cite{zipperle2022provenance}, was developed to support research on APT-focused cybersecurity solutions~\cite{darpa-tc3}. Over two weeks, adversarial teams executed APT-based attacks and documented their activities in ground truth reports~\cite{ground-truth-tc3}. These reports provide summaries of attack stages, key attack indicators, and detailed event logs with timestamps; however, they do not specify the exact malicious system entities involved as ground-truth labels. 
Our evaluation covered two Linux-based hosts (THEIA and TRACE) and one FreeBSD-based host (CADETS). 

\subsubsection{DARPA OpTC}
The DARPA OpTC dataset includes data from 1,000 Windows OS hosts simulating a large enterprise environment~\cite{darpa-optc,anjum2021analyzing}.
It spans seven days, with only benign activity during the first four days. The final three days contain both benign and malicious activity, where a red team conducts APT-style attacks. These attacks cover the full APT lifecycle~\cite{ground-truth-optc}, including initial compromise, internal reconnaissance, command \& control, persistence, and trace-covering actions. 
Following prior studies~\cite{rehman2024flash,aly2024megr}, we focus our evaluation on the three hosts with the highest volume of attack traces, based on the ground truth provided by FLASH. This selection enables fair comparison with existing methods while still presenting a challenging detection setting due to the low proportion of malicious nodes (0.024\%).

\subsubsection{Simulated NODLINK}
This dataset, released by NODLINK~\cite{li2023nodlink}, simulates the internal environment of a security company, Sangfor. Data were collected from three hosts: an Ubuntu 20.04 server, a Windows Server 2012 (WS 12), and a Windows 10 desktop host (W10). The dataset includes attack descriptions and ground-truth labels, which we used in our evaluation. This dataset enabled us to benchmark our system's performance against NODLINK.

\subsection{Evaluation Setup}
We train {\sysName} on benign traces ($D_{b}$) collected from the provenance graph of a specific host. Then, we test {\sysName} using graphs containing both malicious and benign traces, excluding $D_{b}$. 
We ensure fair and consistent evaluation across the baselines (THREATRACE~\cite{wang2022threatrace}, FLASH~\cite{rehman2024flash}, MAGIC~\cite{jia2024magic}, and KAIROS~\cite{jia2024magic}) using the same datasets, labels, and metrics. 
For NODLINK~\cite{li2023nodlink}, reproduction on DARPA TC3 was not possible due to the lack of access to the specific data subset and ground truth labels used in their evaluation. Instead, we relied on the reported TC3 results and reproduced experiments on their simulated dataset using the official system.
We also reproduced some baselines: FLASH results closely matched published ones, while THREATRACE showed high variance (e.g., F1-score $0.595\pm0.434$ on TC3). We contacted the authors, who confirmed this instability. Therefore, we rely on original papers to compare OCR-APT with each method's best-performing version.
We follow the same evaluation setup as prior work~\cite{wang2022threatrace,rehman2024flash}, where true positives are anomalous nodes correctly identified as abnormal or those with 2-hop neighbors flagged as abnormal. False positives are benign nodes mistakenly flagged despite having no anomalous nodes within two hops.

\subsubsection{Parameter Setting}
To optimize detection accuracy, we conducted a hyperparameter tuning experiment to select the default parameters, which were subsequently used in all our evaluations. For our {\gnnModel} models, we implemented three layers of RGCN, utilizing a 32-dimensional embedding vector and a learning rate\footnote{The learning rate determines the step size for parameter updates during training~\cite{yang2020hyperparameter}.} of 0.005. 
The contamination factor was set to range between $Min_{con} = 0.001$ and $Max_{con} = 0.05$. Following prior work~\cite{liu2024pygod}, we set $\beta = 0.5$.
The number of seed nodes $n_{seed}$ for subgraph construction was set to 15, with a maximum of 5000 edges per subgraph ($max_{e}$).
We assess the abnormality levels of the constructed subgraphs as follows. Subgraphs with an anomaly score below 10 are classified as having minor abnormalities. Those with scores between 10 and 100 exhibit moderate abnormalities. Scores between 100 and 1000 indicate significant abnormalities, and scores exceeding 1000 are categorized as critical. In our evaluation, subgraphs with moderate abnormalities or higher are labeled as anomalous.

\subsubsection{Infrastructure} 
Our experiments were conducted on a Linux system equipped with 64 cores and 256 GB of RAM. We developed {\sysName} using Python and Bash scripts, leveraging PyTorch Geometric~\cite{NEURIPS2019_9015} for training GNN models and NetworkX~\cite{networkx} for subgraph construction. Our {\gnnModel} is built on top of the PyGOD~\cite{liu2024pygod} library. 
Provenance graphs are stored in the GraphDB~\cite{graphdb} RDF graph database, which supports the RDF-star format used in our system~\cite{aly2024megr}. We developed our RAG-based pipeline using LlamaIndex~\cite{Liu_LlamaIndex_2022}, which offers a vector store and API calls for various LLMs. For our main LLM, we used GPT-4o-mini~\cite{openai2024gpt4o} with temperature set to 0 to ensure accurate and deterministic results~\cite{ullah2024llms}.
As part of our ablation study, we tested OpenAI's embedding models and selected `text-embedding-3-large'~\cite{openai2024textembedding3} for indexing due to its strong performance.
The entire system was implemented in approximately 5,000 lines of code.

\begin{table}[t]
\centering
\caption{Detection accuracy of {\sysName} in comparison with SOTA anomaly detection systems on DARPA TC3, DARPA OpTC, and Simulated NODLINK datasets.}
\label{tab:all-anomaly-detector}
\resizebox{\columnwidth}{!}{%
\begin{tabular}{ccccc}
\toprule
Dataset & System & Precision & Recall & F1-Score \\ 
\midrule

\multirow{6}{*}{\begin{tabular}[c]{@{}c@{}}TC3\\ (CADETS)\end{tabular}}      
& THREATRACE & 0.90 & 0.99 & 0.95 \\
& MAGIC & 0.94 & 0.99 & 0.97 \\
& KAIROS & 1.00 & 0.95 & 0.97 \\
& NODLINK & 0.14 & 1.00 & 0.25 \\
& FLASH & 0.95 & 0.99 & 0.97 \\
& \textbf{OCR-APT} & \textbf{1.00} & \textbf{1.00} & \textbf{1.00} \\ \hline

\multirow{6}{*}{\begin{tabular}[c]{@{}c@{}}TC3\\ (TRACE)\end{tabular}}    
& THREATRACE & 0.72 & 0.99 & 0.83 \\
& MAGIC & 0.99 & 0.99 & 0.99 \\
& NODLINK & 0.25 & 0.98 & 0.40 \\
& FLASH & 0.95 & 0.99 & 0.97 \\
& \textbf{OCR-APT} & \textbf{1.00} & \textbf{1.00} & \textbf{1.00} \\ \hline

\multirow{6}{*}{\begin{tabular}[c]{@{}c@{}}TC3\\ (THEIA)\end{tabular}}    
& THREATRACE & 0.87 & 0.99 & 0.93 \\
& MAGIC & 0.98 & 0.99 & 0.99 \\
& KAIROS & 1.00 & 0.95 & 0.97 \\
& NODLINK & 0.23 & 1.00 & 0.37 \\
& FLASH & 0.93 & 0.99 & 0.96 \\
& \textbf{OCR-APT} & \textbf{1.00} & \textbf{1.00} & \textbf{1.00} \\ \hline

\multirow{3}{*}{\begin{tabular}[c]{@{}c@{}}OpTC \\ (201)\end{tabular}}      
& THREATRACE & 0.84 & 0.85 & 0.84 \\
& FLASH & 0.90 & 0.92 & 0.91 \\
& \textbf{OCR-APT} & \textbf{1.00} & \textbf{0.88} & \textbf{0.94} \\ \hline

\multirow{3}{*}{\begin{tabular}[c]{@{}c@{}}OpTC\\ (501)\end{tabular}}     
& THREATRACE & 0.85 & 0.87 & 0.86 \\
& FLASH & 0.94 & 0.92 & 0.93 \\
& \textbf{OCR-APT} & \textbf{1.00} & \textbf{1.00} & \textbf{1.00} \\ \hline

\multirow{3}{*}{\begin{tabular}[c]{@{}c@{}}OpTC\\ (51)\end{tabular}}      
& THREATRACE & 0.86 & 0.87 & 0.86 \\
& \textbf{FLASH} & \textbf{0.94} & \textbf{0.92} & \textbf{0.93} \\
& OCR-APT & 0.89 & 0.77 & 0.82 \\ \hline

\multirow{2}{*}{\begin{tabular}[c]{@{}c@{}} NODLINK \\ (Ubuntu)\end{tabular}} 
& NODLINK & 0.04 & 0.38 & 0.07 \\
& \textbf{OCR-APT} & \textbf{0.95} & \textbf{1.00} & \textbf{0.97} \\ \hline

\multirow{2}{*}{\begin{tabular}[c]{@{}c@{}} NODLINK \\ (WS 12)\end{tabular}}  
& NODLINK & 0.10 & 0.84 & 0.17 \\
& \textbf{OCR-APT} & \textbf{0.74} & \textbf{0.93} & \textbf{0.82} \\ \hline

\multirow{2}{*}{\begin{tabular}[c]{@{}c@{}} NODLINK \\ (W10)\end{tabular}} 
& NODLINK & 0.14 & 0.68 & 0.23 \\
& \textbf{OCR-APT} & \textbf{0.95} & \textbf{0.99} & \textbf{0.97} \\ 
\bottomrule
\end{tabular}
}
\end{table}

\subsection{Evaluation of Detection Accuracy}
We compared {\sysName} with state-of-the-art (SOTA) anomaly detection systems across various granularities: nodes (THREATRACE and MAGIC), time windows (KAIROS), and subgraphs (NODLINK and FLASH). To enable unified evaluation, Table~\ref{tab:all-anomaly-detector} reports the detection accuracy of all systems at the node level, where any node within an anomalous time window or subgraph is labeled anomalous. Results for the SOTA systems are drawn from their original papers\footnote{As THREATRACE does not report evaluation results on the OpTC dataset, we present the results produced and reported by FLASH.}.
Overall, {\sysName} achieved comparable or superior accuracy to existing node-level detectors. However, these systems do not support subgraph-level anomaly detection, which is essential for our LLM-based investigator to reconstruct attack reports.

On the DARPA TC3 dataset, {\sysName} achieved higher recall than KAIROS\footnote{KAIROS was also evaluated on the DARPA OpTC dataset, but its results were not reported at the node level, preventing direct comparison.}. KAIROS detects anomalies over 15-minute time windows, each manually labeled based on ground truth. While it attains 100\% recall at the window level, its node-level recall caps at 95\%, likely due to malicious nodes falling outside the labeled windows, which may contain both benign and malicious entities.
In contrast, {\sysName} provides comparable accuracy while operating at the subgraph level rather than fixed time windows. This ensures that anomalous subgraphs consist only of causally connected anomalous nodes, incorporating benign nodes only when they serve as bridges between anomalous events.

Detecting anomalies at the subgraph level improves alert validation and interpretability but is more challenging than node-level detection. 
{\sysName} achieved perfect accuracy across all DARPA TC3 hosts, whereas NODLINK struggled with a high false-positive rate, reporting a maximum precision of just 0.25 on the TRACE host.
NODLINK uses sentence embeddings, which fail to capture the graph structure. This limitation negatively impacts its accuracy, as observed in the NODLINK dataset\footnote{As NODLINK's authors did not provide per-host results on their simulated dataset, we executed it using their public scripts and metrics without modification.}. 
In contrast, {\gnnModel} leverages both graph structure and node behavior, leading to superior performance. {\sysName} consistently outperformed NODLINK, with its lowest F1-score being 0.82 on the simulated WS 12 host. In this case, {\sysName} missed 3 out of 47 malicious entities and produced 13 false positives among 10,860 benign nodes. Due to the small size of the simulated dataset, minor errors had a considerable impact on evaluation metrics. {\sysName} enhances interpretability through subgraph-level detection without compromising accuracy.

Overall, {\sysName} outperformed all detectors across all hosts, except for OpTC 51, where FLASH achieved higher detection accuracy. In that case, the adversary launched a malicious upgrade attack by installing a backdoored version of Notepad-Plus. During the update process, the backdoor connected to the attacker’s server to download both legitimate updates and a malicious binary. This behavior confused the anomaly detection model, which failed to flag the malicious binary. However, our LLM-based attack investigator successfully identified both the malicious binary and the command-and-control server in the generated attack report. 
Furthermore, {\sysName} demonstrates greater robustness than SOTA anomaly detection systems by avoiding reliance on node features that are susceptible to adversarial manipulation.

Robustness to evasion remains a critical factor for anomaly detection systems. A growing concern in this area is mimicry attacks, where adversaries inject benign activities into attack graphs to evade detection while preserving the core malicious behavior. Provenance-based intrusion detection systems that operate at the graph or path level have proven vulnerable to such tactics~\cite{goyal2023sometimes}. However, the same study suggests that focusing on finer-grained —such as nodes, edges, or subgraphs—can mitigate this risk~\cite{goyal2023sometimes}, a strategy that has shown promise in several recent systems~\cite{rehman2024flash,cheng2024kairos,wang2024incorporating}. {\sysName}'s subgraph-level detection naturally aligns with these insights and offers a promising defense against such evasion. As part of future work, we aim to assess its robustness against a broad range of mimicry and evasion techniques~\cite{goyal2023sometimes,mukherjee2023evading}.

\subsection{Ablation Study}
This section evaluates {\gnnModel} against existing GNN-based anomaly detectors using the simulated NODLINK dataset, chosen for its manageable size. Some baselines failed to run on larger datasets (e.g., CADETS from DARPA TC3) due to memory constraints. We also conduct ablation studies to assess the impact of key components and tune hyperparameters for optimal performance. Each experiment is repeated 10 times, and average results are reported.

\subsubsection{The {\gnnModel} Models.}
We developed six variations of {\sysName}: one with our GNN model, and the rest with existing GNN-based anomaly detection models implemented using the PyGOD~\cite{liu2024pygod} library. These models include AnomalyDAE~\cite{fan2020anomalydae}, CONAD~\cite{xu2022contrastive}, CoLA~\cite{liu2021anomaly}, GAE~\cite{kipf2016variational}, and OCGNN~\cite{ocgnn}. Table~\ref{tab:general-detectors} presents a comparison of the detection accuracy and efficiency of {\sysName} using {\gnnModel} model versus general detectors.

Overall, {\gnnModel} consistently outperforms these detectors in accuracy, as it captures edge types when aggregating node embeddings. In contrast, these detectors are designed for homogeneous graphs and do not incorporate edge types. For example, OCGNN is a one-class classification method similar to {\gnnModel}, but it does not capture edge types. As a result, OCGNN suffers from low precision and struggles to differentiate between normal and anomalous nodes.
CoLA, a self-supervised learning method for graph anomaly detection, achieves slightly higher precision than {\gnnModel} in the W10 host. However, its performance is inconsistent, with an F1-score of approximately 0.4 in the other two hosts.
Autoencoding-based methods, such as GAE, CONAD, and AnomalyDAE, exhibit inconsistent detection accuracy. Notably, both CONAD and AnomalyDAE failed to detect any anomalous nodes in the WS12 host.
Besides, these methods are typically memory-intensive, as they scale quadratically with the number of nodes due to the reconstruction of the complete graph adjacency matrix~\cite{liu2022bond}.

\begin{table}[t]
\centering
\caption{Evaluating APT detection accuracy and efficiency of {\sysName} on the Simulated NODLINK dataset using various GNN-based anomaly detection models. {\gnnModel} is our novel GNN-based model.}
\label{tab:general-detectors}
\resizebox{\columnwidth}{!}{%
\begin{tabular}{ccccccc}
\toprule
Host &
  Model &
  Precision &
  Recall &
  F1-Score &
  \begin{tabular}[c]{@{}c@{}}Detection \\ Time (s)\end{tabular} &
  \begin{tabular}[c]{@{}c@{}}Occupied \\ Memory (GB)\end{tabular} \\ 
  \midrule
\multirow{6}{*}{\begin{tabular}[c]{@{}c@{}} Ubuntu\end{tabular}} &
  AnomalyDAE &
  0.24 &
  0.61 &
  0.34 &
  1,411.67 &
  53.88 \\
                                                                            & CONAD           & 0.34          & 1.00          & 0.51          & 1,431.23 & 53.93 \\
                                                                            & \textbf{GAE}    & \textbf{1.00} & \textbf{0.94} & \textbf{0.97} & 928.60   & 46.27 \\
                                                                            & CoLA            & 0.32          & 0.59          & 0.40          & 182.33   & 13.58 \\
                                                                            & OCGNN           & 0.04          & 1.00          & 0.07          & 941.84   & 31.43 \\
                                                                            & \textbf{OCRGCN} & \textbf{0.95} & \textbf{1.00} & \textbf{0.97} & 828.01   & 41.78 \\ \hline
\multirow{6}{*}{\begin{tabular}[c]{@{}c@{}}WS12\end{tabular}} & AnomalyDAE      & 0.00          & 0.00          & 0.00          & 111.16   & 8.26  \\
                                                                            & CONAD           & 0.00          & 0.00          & 0.00          & 124.34   & 9.03  \\
                                                                            & GAE             & 0.26          & 0.93          & 0.40          & 115.13   & 7.47  \\
                                                                            & CoLA            & 0.30          & 0.47          & 0.37          & 31.80    & 7.80  \\
                                                                            & OCGNN           & 0.32          & 0.98          & 0.48          & 108.39   & 7.80  \\
                                                                            & \textbf{OCRGCN} & \textbf{0.74} & \textbf{0.93} & \textbf{0.82} & 30.87    & 10.82 \\ \hline
\multirow{6}{*}{\begin{tabular}[c]{@{}c@{}}W10\end{tabular}}   & AnomalyDAE      & 0.82          & 0.99          & 0.90          & 266.69   & 63.69 \\
                                                                            & CONAD           & 0.82          & 0.99          & 0.90          & 264.55   & 59.45 \\
                                                                            & GAE             & 0.70          & 1.00          & 0.83          & 159.69   & 7.26  \\
                                                                            & \textbf{CoLA}   & \textbf{0.98} & \textbf{0.99} & \textbf{0.99} & 41.43    & 7.59  \\
                                                                            & OCGNN           & 0.75          & 1.00          & 0.86          & 173.54   & 7.69  \\
                                                                            & \textbf{OCRGCN} & \textbf{0.95} & \textbf{0.99} & \textbf{0.97} & 59.21    & 10.48 \\ 
                                                                            \bottomrule
\end{tabular}%
}
\end{table}

\subsubsection{{\sysName} System Components.} 
\label{components-analysis}
We conducted ablation experiments to assess the impact of {\sysName}’s components. 
Four variations of the system were created: the full system, one without behavior-based features (\textit{Without B-Feat}), one without type-specific models (\textit{Without TS-Mod}), and one without subgraph anomaly detection (\textit{Without SG-Det}). Table~\ref{tab:core-components} shows the detection accuracy and efficiency of each variation.

The variation \textit{Without B-Feat} relies on features from prior work, THREATRACE~\cite{wang2022threatrace}, excluding statistics of the node idle phase and normalization techniques. This variant failed to detect any anomalous nodes in the WS12 host and reduced precision in the other two hosts. These results align with our hypothesis that the statistics of node idle periods assist in distinguishing benign nodes from those associated with APT activity. Additionally, our behavior-based features enhance time efficiency due to feature normalization. 

The \textit{Without TS-Mod} variant employs a single {\gnnModel} model for all node types. While this approach improves time and memory efficiency, it significantly compromises detection accuracy. In the Ubuntu host, recall dropped from 100\% to 44\%, while precision in the WS12 host declined sharply from 74\% to 22\%.
These findings indicate that the model struggles to learn the normal behavior of different node types when relying on a single model.
Training multiple {\gnnModel} models—one per node type—assists in capturing variations in benign behavior, as normal behavior patterns differ across node types. We acknowledge that benign traffic may exhibit multiple behavior patterns; therefore, exploring clustering-based methods may offer a promising direction for future work.

In Table~\ref{tab:core-components}, the \textit{Without SG-Det} variant performs only node-level detection, avoiding the time overhead of subgraph construction and detection. This variant reduces precision across all hosts, underscoring the value of our subgraph anomaly detection in filtering false positives. The results show that subgraph detection improves precision while maintaining high recall. {\sysName}'s subgraph construction reduces false positives by filtering out subgraphs with low abnormality scores. Even if a node's anomaly score is inaccurate, it will not trigger an alert unless it belongs to a highly abnormal subgraph.
These findings highlight the importance of each core component in enhancing {\sysName}'s effectiveness.

\begin{table}[t]
\caption{Comparison of APT Detection accuracy and efficiency of {\sysName} on Simulated NODLINK dataset with all system components, and without the behavior-based features (B-Feat), the type-specific models (TS-Mod), and the subgraph anomaly detection (SG-Det).}
\centering
\label{tab:core-components}
\resizebox{\columnwidth}{!}{%
\begin{tabular}{ccccccc}
\toprule
Host &
  Version &
  Precision &
  Recall &
  F1-Score &
  \begin{tabular}[c]{@{}c@{}}Detection \\ Time (s)\end{tabular} &
  \begin{tabular}[c]{@{}c@{}}Occupied \\ Memory (GB)\end{tabular} \\ \midrule
\multirow{4}{*}{\begin{tabular}[c]{@{}c@{}}Ubuntu\end{tabular}} & Without B-Feat   & 0.41          & 1.00          & 0.58          & 1,380.69 & 52.83 \\
                                                                             & Without TS-Mod   & 1.00          & 0.44          & 0.62          & 24.01    & 18.97 \\
                                                                             & Without SG-Det   & 0.58          & 1.00          & 0.73          & 34.89    & 19.61 \\
                                                                             & \textbf{OCR-APT} & \textbf{0.95} & \textbf{1.00} & \textbf{0.97} & 828.01   & 41.78 \\ \hline
\multirow{4}{*}{\begin{tabular}[c]{@{}c@{}}WS12\end{tabular}}  & Without B-Feat   & 0.00          & 0.00          & 0.00          & 57.02    & 10.65 \\
                                                                             & Without TS-Mod   & 0.22          & 0.93          & 0.36          & 12.37    & 10.47 \\
                                                                             & Without SG-Det   & 0.32          & 1.00          & 0.48          & 22.58    & 10.82 \\
                                                                             & \textbf{OCR-APT} & \textbf{0.74} & \textbf{0.93} & \textbf{0.82} & 30.87    & 10.82 \\ \hline
\multirow{4}{*}{\begin{tabular}[c]{@{}c@{}}W10\end{tabular}}    & Without B-Feat   & 0.89          & 0.99          & 0.94          & 134.26   & 10.30 \\
                                                                             & Without TS-Mod   & 0.80          & 0.99          & 0.89          & 78.74    & 10.10 \\
                                                                             & Without SG-Det   & 0.80          & 1.00          & 0.89          & 20.28    & 10.48 \\
                                                                             & \textbf{OCR-APT} & \textbf{0.95} & \textbf{0.99} & \textbf{0.97} & 59.21    & 10.48 \\ \bottomrule
\end{tabular}%
}
\end{table}

\subsubsection{Hyperparameter Tuning.}
\label{hyperparameter}
To optimize the F1-score, hyperparameter tuning was performed using Bash scripts to systematically evaluate a range of parameter configurations. For GNN model training, this included variations in the number of RGCN layers \{2, 3, 4\}, graph embedding vector sizes \{32, 64, 92\}, and learning rates \{0.005, 0.001, 0.0005\}. 
We also varied the $\beta$ parameter of the one-class SVM in \{0.3, 0.4, 0.5, 0.6, 0.7\} and observed minimal impact on performance; thus, we used PyGOD's~\cite{liu2024pygod} default value ($\beta = 0.5$).
For subgraph construction, we considered parameters such as the number of seed nodes \{10, 15, 20\} and the maximum edges per subgraph \{5000, 10000\}. 
We also evaluated the impact of using two-hop expansion during subgraph construction. While two-hop expansion slightly reduces false negatives, it significantly increases false positives. For example, in THEIA, false negatives reduced from 5 to 2, but false positives rose sharply from 0 to 21.5 K, reducing precision from 1.0 to 0.5. Other datasets showed minimal impact (see Appendix~\ref{hop-expansion} for more results). Based on these experiments, we adopt one-hop expansion as the default configuration for subgraph construction.
While the selected default parameters were applied consistently across all hosts, future datasets with varying levels of complexity may benefit from dataset-specific hyperparameter tuning to ensure optimal performance.

\subsection{Evaluation of Recovered Attack Reports}
We evaluated the quality of our LLM-based attack investigator by comparing our recovered attack reports to the ground truth reports provided by DARPA~\cite{ground-truth-tc3, ground-truth-optc}.  
Table~\ref{tab:llm-results-summary} provides a summary of the detected IOCs and APT stages across all reports. The APT stages include Initial Compromise (IC), Internal Reconnaissance (IR), Command and Control (C\&C), Privilege Escalation (PE), Lateral Movement (LM), Maintain Persistence (MP), Data Exfiltration (DE), and Covering Tracks (CT).
In most cases, our recovered reports covered the majority of APT stages (highlighted in green). They clearly specify artifacts, such as command-and-control servers, malicious executable files, and exploited processes involved in the attacks.
For example, reports recovered from the TRACE host captures all performed attack stages, including initial compromise, persistence, command and control, and internal reconnaissance. 
Simplified versions of the recovered reports\footnote{Full versions of the reports are available at \url{https://github.com/CoDS-GCS/OCR-APT/tree/main/recovered_reports}} are provided in Appendix~\ref{appendix-reports}. 
Missed stages (highlighted in red) in DARPA TC3 dataset primarily resulted from OS log parsing issues. This led to the missing of process attributes on the CADETS host and file attributes on the THEIA host. 

For the DARPA OpTC dataset, the recovered reports captured most APT stages but struggled to identify the lateral movement and initial compromise phases. Detecting lateral movement was beyond the scope of this work, as {\sysName} does not process network traffic logs. In future work, we plan to address this limitation by integrating network traffic analysis with specialized detectors.
The initial compromise stage was challenging to detect because the initial payload files remained inactive during the attack. As a result, the anomaly detection models did not flag them as suspicious. However, further analysis revealed that these overlooked artifacts were directly connected to detected IOCs. 
To mitigate missing IOCs, our approach enriches reports with subgraphs surrounding key IOCs. The LLM-based investigator explores these anomalous subgraphs to uncover related artifacts missed by the detection model. For example, on the OpTC 51 host, it identified a malicious binary and a C\&C IP that had been initially overlooked. This enrichment helps the LLM infer additional threats and improves overall detection.
Despite these limitations, the recovered reports provide clear and detailed accounts of the attack scenarios. They align well with the attack timestamps from the ground truth and reference most key artifacts. Moreover, the reports are written in a human-like narrative style, similar to CTI reports.

\begin{table}[t]
\caption{Evaluation of recovered attack reports using both commercial (GPT-4o-mini) and local (LLAMA3-8B) LLMs on DARPA TC3 and OpTC datasets. The table shows the number of detected IOCs and APT attack stages, with total counts in parentheses. Detected stages are highlighted in green, while missed stages are shown in red.}
\centering
\label{tab:llm-results-summary}
\resizebox{\columnwidth}{!}{%
\begin{tabular}{|c|ccccc|}
\hline
LLM &
  Dataset & 
  Host & 
  \begin{tabular}[c]{@{}c@{}}\# Detected \\ IOCs\end{tabular} &
  \begin{tabular}[c]{@{}c@{}}\# Detected \\ APT Stages\end{tabular} &
  Detected APT Stages \\ \hline
 &                                                                        & CADETS & 11 (16) & 5 (6) & \textcolor{red}{IC}{\color[HTML]{38761D}, MP, PE, C\&C, IR, CT}         \\
 &                                                                        & TRACE  & 6 (7)   & 4 (4) & \color[HTML]{38761D} {IC, MP, C\&C, IR}                 \\
 & \multirow{-3}{*}{\begin{tabular}[c]{@{}c@{}}DARPA \\ TC3\end{tabular}} & THEIA  & 5 (7)   & 5 (6) & \color[HTML]{38761D}{IC, MP, PE, C\&C, IR,} \textcolor{red}{CT}             \\ \cline{2-6} 
 &                                                                        & 201    & 5 (6)   & 5 (7) & {\color[HTML]{38761D} \textcolor{red}{IC}, MP, PE, C\&C, IR, \textcolor{red}{LM} , CT}    \\
 &                                                                        & 501    & 7 (11)  & 5 (8) & {\color[HTML]{38761D} \textcolor{red}{IC}, MP, PE, C\&C, IR, \textcolor{red}{LM, DE}, CT} \\
\multirow{-6}{*}{GPT-4o-mini} &
  \multirow{-3}{*}{\begin{tabular}[c]{@{}c@{}}DARPA \\ OpTC\end{tabular}} &
  51 &
  8 (10) &
  4 (6) &
  \textcolor{red}{IC} {\color[HTML]{38761D} , MP, PE, C\&C, IR, \textcolor{red}{LM}} \\ \hline
 &                                                                        & CADETS & 10 (16) & 5 (6) & \textcolor{red}{IC}{\color[HTML]{38761D}, MP, PE, C\&C, IR, CT}         \\
 &                                                                        & TRACE  & 5 (7)   & 4 (4) & \color[HTML]{38761D} {IC, MP, C\&C, IR}                 \\
 & \multirow{-3}{*}{\begin{tabular}[c]{@{}c@{}}DARPA \\ TC3\end{tabular}} & THEIA  & 5 (7)   & 5 (6) & \color[HTML]{38761D}{IC, MP, PE, C\&C, IR,} \textcolor{red}{CT}             \\ \cline{2-6} 
 &                                                                        & 201    & 2 (6)   & 3 (7) & {\color[HTML]{38761D} \textcolor{red}{IC}, MP, PE, \textcolor{red}{C\&C, IR, LM} , CT}    \\
 &                                                                        & 501    & 7 (11)  & 5 (8) & {\color[HTML]{38761D} \textcolor{red}{IC}, MP, PE, C\&C, IR, \textcolor{red}{LM, DE}, CT} \\
\multirow{-6}{*}{LLAMA3-8B} &
  \multirow{-3}{*}{\begin{tabular}[c]{@{}c@{}}DARPA \\ OpTC\end{tabular}} &
  51 &
  7 (10) &
  4 (6) &
  \textcolor{red}{IC} {\color[HTML]{38761D} , MP, PE, C\&C, IR, \textcolor{red}{LM}} \\ \hline
\end{tabular}%
}
\end{table} 

To evaluate our system in a practical setting, we conducted experiments using locally deployed open-source LLMs. These models preserve data privacy by eliminating the need to transmit sensitive system logs to commercial providers, offering a cost-effective solution for long-term use.
We ran the pipeline on a local machine with 4 CPU cores, an 8GB GPU, and 22GB of RAM. 
An ablation study guided the selection of the most effective local LLM and embedding model. We evaluated six local LLMs and seven embedding models. The best setup—LLaMA3 (8B)\cite{llama3modelcard} paired with IBM's open-source `granite-embedding-125m-english'\cite{granite2024embedding}—achieved comparable performance to ChatGPT. As shown in Table~\ref{tab:llm-results-summary}, this configuration detected the same APT stages as ChatGPT on all hosts except DARPA OpTC 501, where it missed two stages.
While the quality of the comprehensive reports declined, the overall investigation results remained reliable and informative.
This strong performance demonstrates the effectiveness of our pipeline, even when using lightweight, locally deployed models. LLAMA3’s relatively small size further suggests that {\sysName}’s performance is not driven by memorization of benchmark datasets.

Furthermore, {\sysName} systematically validates generated reports against detected anomalies, ensuring that outputs are grounded in actual data rather than relying on prior model knowledge or LLM hallucinations. This is achieved by modularizing the investigation into subtasks, each guided by specialized Chain-of-Thought (CoT) prompts.
To assess the impact of this design, we compared it against a baseline that uses a single CoT-based prompt to generate the entire attack report, skipping intermediate steps like IOC extraction and validation. The baseline produced fewer detected APT stages and, more critically, frequent hallucinations—including fabricated entities like \texttt{malicious.exe}, \texttt{suspicious\_process.exe}, and \texttt{vulnerable\_service.exe}, which were not present in the source provenance graph. These results highlight the advantages of our modular pipeline and its integrated validation mechanism.

Our {\sysName} automatically analyzes audit logs and generates valuable insights in the form of human-like security reports. These reports cover most APT stages and include key IOCs. Hence, they provide a clear overview of the attack progression and highlight critical indicators. These reports save security analysts significant time and enable them to quickly identify key patterns. This leads to more focused and efficient investigations.

\subsection{Discussion and Limitations}
\begin{table}[t]
\centering
\caption{Detection accuracy of {\sysName} and FLASH on DARPA TC3 without neighbor-based assumptions (original metric results in brackets).}
\label{tab:general-evaluation}
\resizebox{\columnwidth}{!}{%
\begin{tabular}{ccccc}
\hline
Dataset                 & System           & Precision            & Recall               & F1-Score             \\ \hline
\multirow{2}{*}{CADETS} & FLASH            & 0.65 (0.93)          & 1.00 (1.00)          & 0.79 (0.96)          \\
                        & \textbf{OCR-APT} & \textbf{1.00 (1.00)} & \textbf{1.00 (1.00)} & \textbf{1.00 (1.00)} \\ \hline
\multirow{2}{*}{TRACE}  & FLASH            & 0.66 (0.95)          & 0.99 (0.99)          & 0.79 (0.97)          \\
                        & \textbf{OCR-APT} & \textbf{0.87 (1.00)} & \textbf{1.00 (1.00)} & \textbf{0.93 (1.00)} \\ \hline
\multirow{2}{*}{THEIA}  & FLASH            & 0.72 (0.92)          & 0.99 (1.00)          & 0.84 (0.96)          \\
                        & \textbf{OCR-APT} & \textbf{0.98 (1.00)} & \textbf{1.00 (1.00)} & \textbf{0.99 (1.00)} \\ \hline
\end{tabular}%
}
\end{table}
\subsubsection{Evaluation Metric.}
We follow the evaluation setup used in prior work~\cite{wang2022threatrace,rehman2024flash,jia2024magic,cheng2024kairos}, which treats neighboring nodes of compromised ones as part of the attack. Although this assumption may not always hold~\cite{cheng2024kairos}, it ensures consistency with existing system evaluations.
To further assess {\sysName}, we conducted additional experiments using a stricter metric that considers only directly identified malicious nodes as true positives, without relying on neighbor-based assumptions.
Table~\ref{tab:general-evaluation} presents the detection results of {\sysName} and FLASH (using FLASH's official implementation) under this setting.
The results show that {\sysName} maintains high precision and recall, with only minor precision drops in some cases. In contrast, FLASH exhibits a significant decline in precision. For instance, on the CADETS host, FLASH's precision dropped from 0.93 to 0.65, while {\sysName} remained stable. On TRACE, FLASH fell from 0.95 to 0.66, whereas {\sysName} declined slightly to 0.87. On THEIA, FLASH dropped from 0.92 to 0.72, while {\sysName} maintained a high precision of 0.98.
These results highlight {\sysName}'s reliability under stricter evaluation and its advantage over existing baselines.
A broader assessment of evaluation metrics is a promising direction for future work toward establishing best practices in anomaly detection benchmarking.
Though narrative clarity remains challenging to quantify, {\sysName}'s structured reports offer more interpretable outputs than prior methods, encouraging future efforts to formalize this aspect.

\subsubsection{Multiple Attack Handling.}
One limitation of our approach lies in the subgraph construction process, which may inadvertently merge multiple attacks into a single subgraph when they share system entities (e.g., processes). While this can be useful for capturing shared infrastructure or correlated activity, it may also obscure the boundaries between causally unrelated attacks, potentially confusing the investigation reports. Although our system can manage causally disconnected attacks to some extent, accurately distinguishing them within a shared subgraph remains challenging. Future work could address this by segmenting subgraphs based on behavioral signatures or by enhancing the LLM investigation module to better identify boundaries between separate attacks.

\subsubsection{Model Generalization.}
The model's ability to generalize is influenced by the extent to which benign behavior is represented in the training data---a known limitation of anomaly detection. Our approach mitigates this by incorporating structural and behavioral features that support generalization, as reflected in the consistently high precision observed across different hosts. However, unseen benign patterns can still lead to false positives. To address this, future work could investigate model adaptation strategies that incorporate analyst feedback through semi-supervised learning.

\section{Related Work}
\label{related-work}
Recent research on provenance-based APT detection~\cite{zipperle2022provenance} can be categorized into two main approaches: heuristic-based and anomaly-based methods~\cite{inam2022sok}. 
In Section~\ref{background}, we discussed the limitations of anomaly detection systems. This section complements the discussion on related work by focusing on heuristic techniques and the emerging role of LLMs in cybersecurity, highlighting how {\sysName} differs from existing work.

\paragraph{LLMs in Cybersecurity.}
LLMs have been applied across diverse cybersecurity tasks, including software vulnerability detection~\cite{lu2024grace, ullah2024llms, ndss2025mammoth}, fuzzing~\cite{oliinyk2024fuzzing}, automated patching~\cite{kulsum2024case, pearce2023examining}, threat detection (e.g., DDoS and phishing)~\cite{guastalla2023application, li2024dollm, li2024knowphish}, penetration testing~\cite{deng2024pentestgpt}, and malware reverse engineering~\cite{hu2024degpt}. In threat intelligence, LLMs help extract knowledge graphs from CTI reports~\cite{cheng2024ctinexus, fieblinger2024actionable, zhang2025attackg}, with benchmarks like AttackSeqBench~\cite{yong2025attackseqbench} assessing LLM effectiveness.
The potential of LLMs for anomaly detection has been explored in a recent survey~\cite{cheng2025sok}.
In contrast, {\sysName} uniquely applies LLMs to reconstruct APT stories from anomalous subgraph alerts. By combining subgraph-level anomaly detection with LLM-driven tasks, such as IOC extraction, stage identification, and report generation, {\sysName} produces interpretable and context-rich reports.

\paragraph{Heuristic-based Detection.}
These systems identify malicious behavior through rules, graph matching, or supervised learning on known attacks. Rule-based approaches~\cite{milajerdi2019holmes, hossain2017sleuth, hassan2020tactical, hossain2020combating} rely on expert-defined specifications derived from TTPs, but they are prone to high false positives or miss zero-day threats~\cite{wang2022threatrace}. CAPTAIN~\cite{wang2024incorporating} improves this by tuning rules with benign data.
Graph matching systems~\cite{altinisik2023provg, poirot, aly2024megr} compare suspicious subgraphs to predefined query graphs derived from CTI reports. While automated graph construction is possible~\cite{poirot}, these systems struggle with novel behaviors not covered in the queries. Similarly, supervised models~\cite{chen2022apt, yan2022deepro} trained on labeled datasets are limited by the scarcity and cost of real APT data. APT-KGL~\cite{chen2022apt} augments training data by mining TTPs and CTI reports but still lacks generalization to unseen threats.
Though some systems incorporate Relational GCNs (RGCNs)~\cite{aly2024megr, chen2022apt}, they typically focus on rule-based or supervised learning paradigms. 
In contrast, {\sysName} adopts a fully anomaly-based approach, detecting deviations from normal behavior at a fine-grained subgraph level—enabling identification of both known and unknown APTs.

In summary, {\sysName}'s core innovation lies in integrating subgraph-level anomaly detection with LLM-based attack reconstruction. This approach avoids reliance on static rules or labeled attack data, offering greater adaptability to emerging threats. 
By converting anomaly alerts into detailed, human-readable reports, {\sysName} enhances both detection and interpretability. This makes {\sysName} a robust and versatile solution for APT defense.

\section{Conclusion}
\label{conclusion}

We proposed {\sysName}, a system that automatically detects APTs and recovers attack reports from provenance graphs. We developed {\sysName} based on our novel GNN-based subgraph anomaly detection and LLM-based investigation. Hence, {\sysName} overcomes the limitations of existing systems. The LLM-based attack investigator generates concise and human-like reports that help analysts efficiently assess and prioritize threats.
Comprehensive evaluations on the DARPA TC3, OpTC, and NODLINK datasets show that {\sysName} consistently outperforms state-of-the-art subgraph anomaly detection systems in detection accuracy. It also enhances the interpretability of results. Additionally, the ablation study demonstrates that {\sysName} effectively balances detection accuracy with memory and time efficiency.
By integrating GNN-based detection with LLM-guided interpretation, {\sysName} significantly advances APT detection and streamlines alert verification, bridging the gap between low-level telemetry and high-level analyst insight.

\begin{acks}
This research is funded by NSERC-CSE Research Communities Grant. 
Researchers funded through the NSERC-CSE Research Communities Grants do not represent the Communications Security Establishment Canada or the Government of Canada. Any research, opinions or positions they produce as part of this initiative do not represent the official views of the Government of Canada. 
\end{acks}
 
\bibliographystyle{ACM-Reference-Format}
\bibliography{references}

\appendix
\section{Provenance Graphs Schema}
\label{appendix-datasets}
\begin{figure}[t]
  \centering
  \includegraphics[width = \columnwidth]{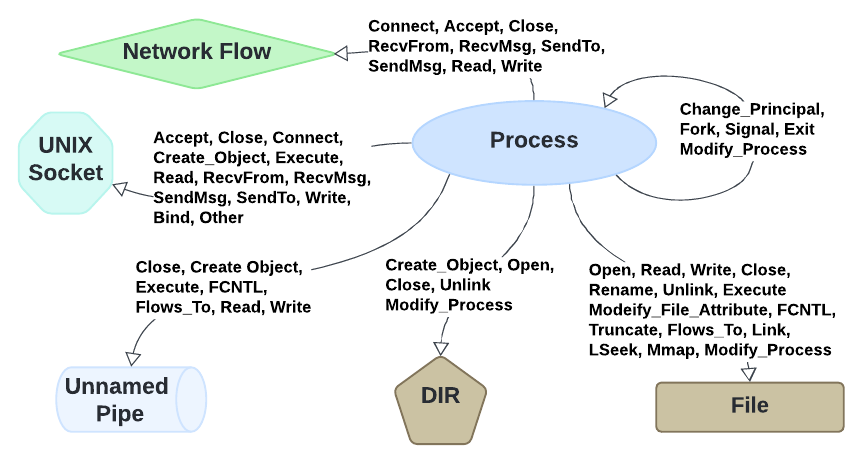}
  \caption{Schema of provenance graphs for the CADETS host.}
  \label{fig:schema}
\end{figure}

Figure~\ref{fig:schema} presents the schema of the provenance graph for the CADETS host, illustrating system entities and the events connecting them. 
The schema includes a diverse range of system entities such as processes, files, and network flows. 
The relationships between these entities include actions such as `read', `write', and `execute', as well as network communications like `send' and `receive'.

\section{LLM Prompts}
\label{llm-prompts}
In this section, we present the prompts and instructions used by our LLM-based attack investigation module.
The system begins by configuring the LLM as an APT investigator, responsible for producing factual and well-structured attack reports.

\begin{center}
\doublebox{
  \begin{minipage}{0.95\columnwidth}
  \textbf{Investigator Instructions:}
  You are an advanced persistent threat (APT) attack investigator, skilled at summarizing log events related to anomaly detection alerts into comprehensive attack reports. You possess deep expertise in APTs, Cyber Threat Intelligence (CTI), and operating system security. \\
  Guidelines:  
    Focus on delivering factual, high-quality analysis in a human-like narrative. 
    Ensure all information is accurate and directly sourced from the document. Do not introduce any details not present in the document, avoiding any fabrications or hallucinations. 
    Keep a detailed account of the attack's execution, including specific timestamps. 
    All responses should be formatted in Markdown. \\
    Definitions: 
    The APT stages are: Initial Compromise, Internal Reconnaissance, Command and Control, Privilege Escalation, Lateral Movement, Maintain Persistence, Data Exfiltration, and Covering Tracks. 
    Indicators of Compromise (IOCs) include: External IP addresses. Suspicious or executable files suspected to be potential threats. Processes with moderate to high likelihood of exploitation. \\
    Your task is to generate an attack report that includes the following sections: 
    A concise summary of the attack behavior, detailing key events and actions taken during the incident. Where applicable, specify the corresponding stage of the APT attack. 
    A table of IOCs detected in the document. Based on your cybersecurity expertise, add a concise security context beside each detected IOC, including the legitimate usage and exploitation likelihood.
    A list of chronological log of actions, organized by minute.
  \end{minipage}
}
\end{center}

The investigation workflow is broken down into a sequence of subtasks, each guided by a specialized prompt.
For each anomalous subgraph, the LLM is first prompted to extract a list of relevant IOCs, returned in Python list format.

\begin{center}
\doublebox{
\begin{minipage}{0.95\columnwidth}
    \textbf{Retrieve IOCs Prompt:}
    The provided document contains log events related to anomaly detection alerts. 
    Extract the list of IOCs from the document $ASG_{doc}$. 
    Return the output only as a Python list, formatted as: [`IOC1', `IOC2', `IOC3', etc].
\end{minipage}
}
\end{center}

The LLM then uses the IOC list to summarize the serialized subgraph into a structured attack report.  

\begin{center}
\doublebox{
\begin{minipage}{0.95\columnwidth}
    \textbf{Summarize Report Prompt:}
    Based on the logs in document $ASG_{doc}$ and the extracted IOCs list: [$IOC_{\text{lst}}$]. 
     Summarize the $ASG_{doc}$ document into an attack report. \\
     The attack report includes the following sections: 
     A concise summary of the attack behavior, detailing key events and actions taken during the incident. Where applicable, specify the corresponding stage of the APT attack.
     A table of IOCs detected in the document. Based on your cybersecurity expertise, add a concise security context beside each detected IOC, including the legitimate usage and exploitation likelihood.
     A list of chronological log of actions, organized by minute.
\end{minipage}
}
\end{center}

After that, the system prompts the LLM to extract the top IOCs associated with each APT stage from attack reports.

\begin{center}
\doublebox{
\begin{minipage}{0.95\columnwidth}
    \textbf{Retrieve IOCs per APT stage Prompt:}
    The provided reports names are: [$R_{\text{atk}}$].
    Extract the three highest-priority IOCs related to the stage: $stg$ from each provided reports. 
    Focus on external IP addresses, suspicious or executable files, malicious processes, and exploitable processes. 
    Return the output only as a Python list, formatted as: [`IOC1', `IOC2', `IOC3', etc].
\end{minipage}
}
\end{center}

The LLM then compiles all attack reports and IOCs into a comprehensive attack report.

\begin{center}
\doublebox{
\begin{minipage}{0.95\columnwidth}
    \textbf{Summarize Comprehensive Report Prompt:}
    Based on the provided reports and the extracted IOCs list: [{$IOC_{\text{lst}}$}]. 
    Summarize all provided reports into a comprehensive attack report. 
    Consider all external IP addresses, suspicious or executable files, malicious processes, and exploitable processes referenced in the provided reports.
\end{minipage}
}
\end{center}

Next, the system initializes a second LLM as a judge, with a role-specific instruction set for a security analyst, who prioritizes IOCs for deeper inspection.

\begin{center}
\doublebox{
  \begin{minipage}{0.95\columnwidth}
  \textbf{Analyst Judge Instructions:}
  You are a highly skilled security analyst specializing in Advanced Persistent Threats (APTs), Cyber Threat Intelligence (CTI), and operating system security. Your expertise includes reviewing attack reports and providing actionable insights. \\
  The APT attack stages are: Initial Compromise, Internal Reconnaissance, Command and Control, Privilege Escalation,  Lateral Movement, Maintain Persistence, Data Exfiltration, and Covering Tracks. \\
  Your task is to analyze the provided attack report and identify key Indicators of Compromise (IOCs) for further investigation. IOCs include external IP addresses, processes with moderate to high exploitation likelihood, and associated suspected files. 
  Focus on identifying IOCs whose contextual analysis could uncover additional APT attack stages, enabling a comprehensive understanding of the full attack scenario. 
  Prioritize IOCs directly tied to malicious activity, such as command-and-control IPs or malicious executable binaries, while deprioritizing general system processes or indicators linked to benign activities. 
  \end{minipage}
}
\end{center}

The judge LLM is prompted to select the highest-priority IOC in the comprehensive report to guide further investigation.

\begin{center}
\doublebox{
\begin{minipage}{0.95\columnwidth}
    \textbf{Select IOC by LLM Judge Prompt:}
    Review the attack report to identify the highest-priority $node_{\text{type}}$ IOC for further investigation, that could aid in uncovering additional APT attack stages. 
    Return the IOC only, formatted as `IOC'.
\end{minipage}
}
\end{center}

The selected IOC is used to query historical context from the graph database, which is summarized into additional reports. The final step prompts the LLM to enrich the comprehensive report using this additional context.

\begin{center}
\doublebox{
\begin{minipage}{0.95\columnwidth}
    \textbf{Enrich Comprehensive Report Prompt:}
    Enrich the comprehensive attack report $R_{\text{comp}}$ by incorporating the summary of the attack report $R_{\text{atk}}$. 
    Consider all external IP addresses, suspicious or executable files, malicious processes, and exploitable processes referenced in the provided reports.
\end{minipage}
}
\end{center}

\section{Recovered Attack Reports}
\label{appendix-reports}
In this section, we present simplified versions of recovered reports. 

\begin{figure}[t]
  \centering
  \begin{framed}
  \includegraphics[width=\columnwidth]{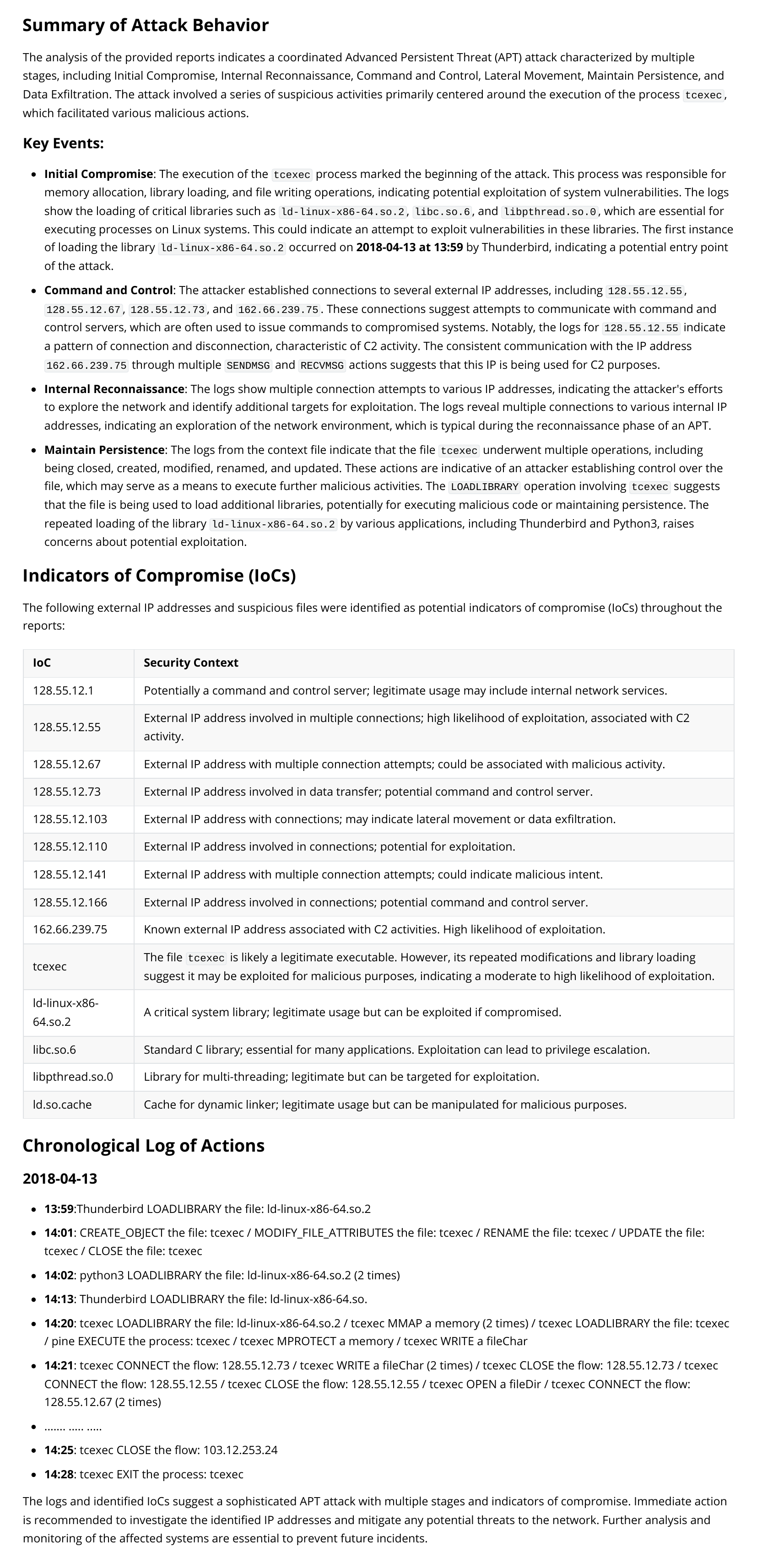}
  \end{framed}
  \caption{A simplified version of the comprehensive attack report recovered from the `TRACE' host in DARPA TC3 dataset.}
  \label{fig:attack_report_trace}
\end{figure}

Figure~\ref{fig:attack_report_trace} shows the comprehensive attack report recovered from the TRACE host in DARPA TC3 dataset. The report details the initial compromise stage, where the attacker leveraged IP \texttt{128.55.12.73} to deliver a malicious executable as an email attachment through the \texttt{thunderbird} process. It identifies the executable file \texttt{tcexec}, which was downloaded to disk, renamed, and had its attributes modified to maintain persistence. Moreover, it captures the use of the server \texttt{162.66.239.75} for command-and-control activities. 

\begin{figure}[t]
  \centering
  \begin{framed}
  \includegraphics[width=\columnwidth]{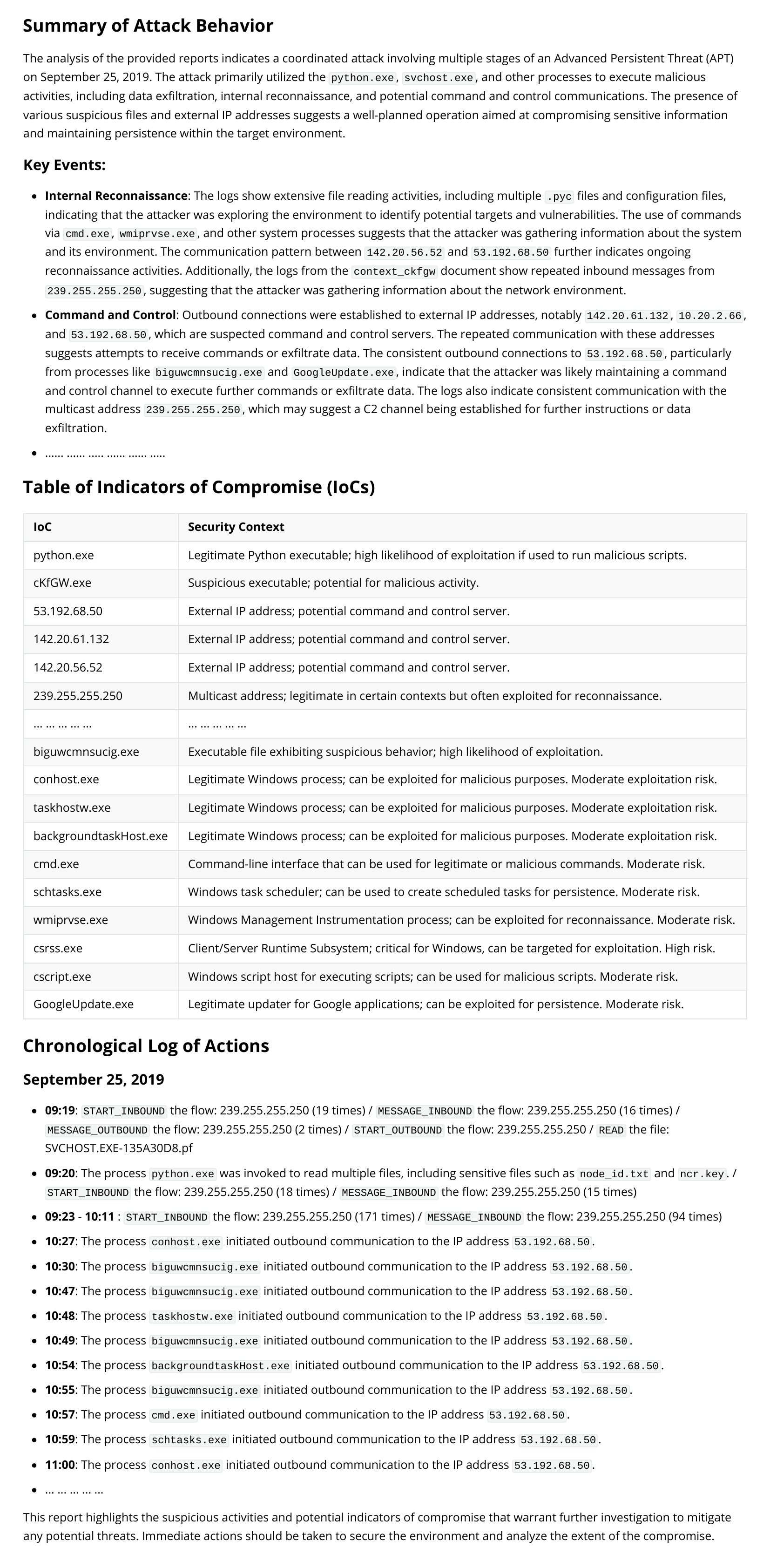}
  \end{framed}
  \caption{A simplified version of the comprehensive attack report recovered from host `51' in the DARPA OpTC dataset, where the red team performed a Malicious Upgrade attack.}
  \label{fig:attack_report_51}
\end{figure}

Figure~\ref{fig:attack_report_51} presents the attack report recovered from host `51' in the DARPA OpTC dataset, detailing the key stages of a malicious upgrade attack.
The attack involved the delivery of the malicious payload \texttt{ckfgw.exe} through a compromised Notepad Plus upgrade process. The recovered report accurately identified several key artifacts, including the malicious binary \texttt{ckfgw.exe}, the command and control server \texttt{53.192.68.50}, the shell script \texttt{cmd.exe} used to execute commands, and the scanned IP address \texttt{142.20.56.52}. It also captured \texttt{schtasks.exe}, which was employed to establish persistence via scheduled tasks. Notably, all these artifacts, along with their timestamps, align with the details provided in the ground truth report. Additionally, the recovered report detected an additional executable binary, \texttt{biGuWCmNsuCIG.exe}, written by \texttt{cKfGW.exe} but not explicitly mentioned in the ground truth report.

\begin{figure}[t]
  \centering
  \begin{framed}
  \includegraphics[width=\columnwidth]{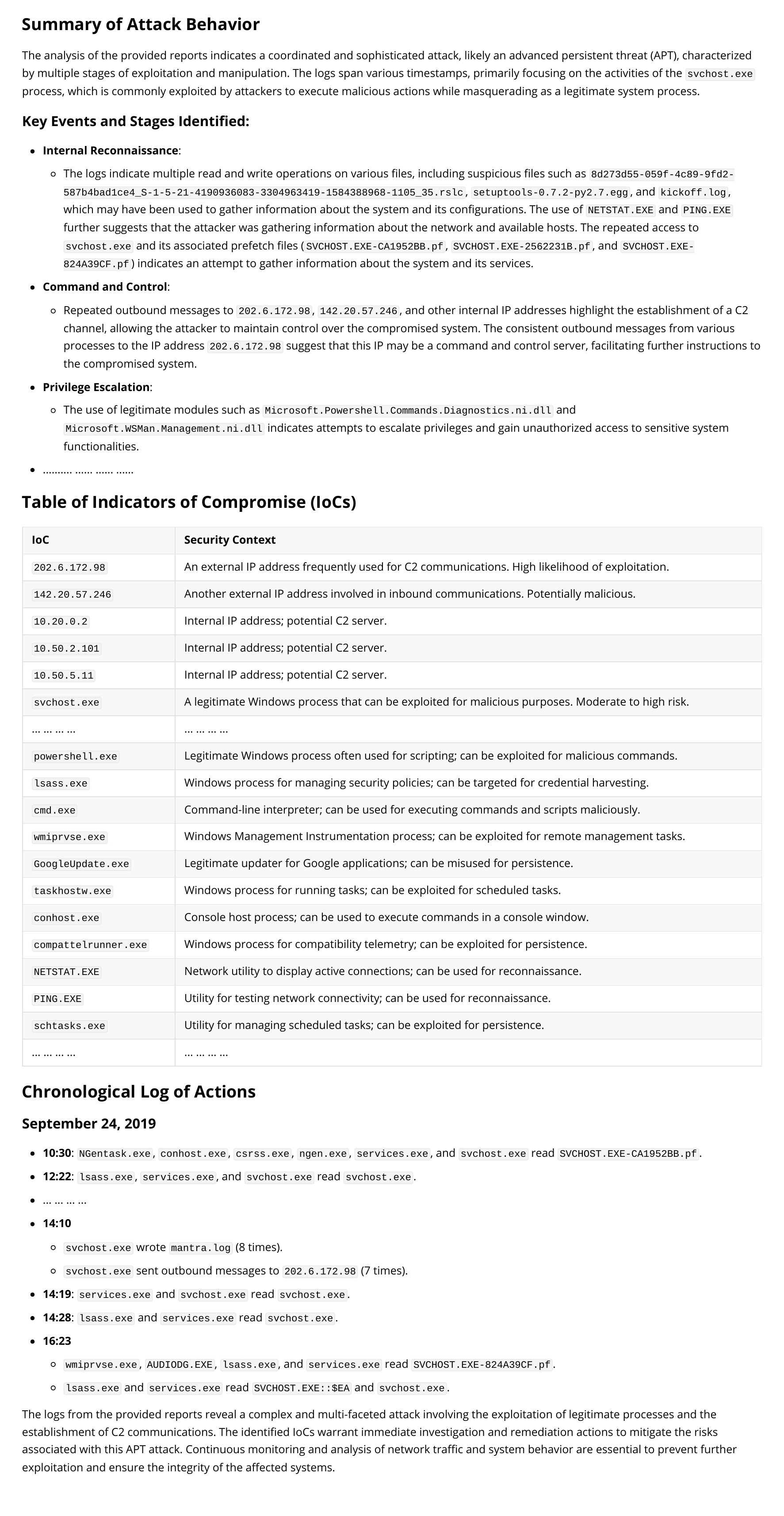}
  \end{framed}
  \caption{A simplified version of the comprehensive report recovered from host `501' in DARPA OpTC dataset, where the red team performed a Custom PowerShell Empire attack.}
  \label{fig:attack_report_501}
\end{figure}

Figure~\ref{fig:attack_report_501} showcases the attack report recovered from host `501' of the DARPA OpTC dataset, highlighting a Powershell Empire attack scenario.
The report successfully identified key elements of the attack, including the \texttt{powershell.exe} script injected during the initial compromise, the command and control server at \texttt{202.6.172.98}, the windows management instrumentation process \texttt{wmiprvse.exe} exploited for privilege escalation, and the \texttt{schtasks.exe} process utilized to manage and automate scheduled tasks. Furthermore, it highlighted the network utility process \texttt{netstat.exe} used for reconnaissance and the scanned IP address \texttt{142.20.57.246}.

\section{Multi-Hop Expansions Analysis}
\label{hop-expansion}
Table~\ref{tab:hop-expansion} compares one-hop and two-hop expansion strategies across multiple datasets. While two-hop expansion slightly reduces false negatives in some cases (e.g., TC3 THEIA, OpTC 201, and OpTC 51), it often introduces a large number of false positives, significantly reducing precision. For example, in THEIA, false negatives reduced from 5 to 2, but false positives rose sharply from 0 to 21.5 K, reducing precision from 1.0 to 0.5. In most datasets, the gains from two-hop expansion are minimal or negligible. Based on this trade-off, we selected one-hop expansion as the default expansion method.

\begin{table}[t]
\caption{Impact of one-hop vs. two-hop expansion on detection performance across datasets.}
\label{tab:hop-expansion}
\centering
\resizebox{\columnwidth}{!}{%
\begin{tabular}{ccccccc}
\hline
Dataset                                                                      & Expansion & FP    & FN & Precision & Recall & F1-Score \\ \hline
\multirow{2}{*}{\begin{tabular}[c]{@{}c@{}}TC3 \\ (CADETS)\end{tabular}}       & one-hop   & 0     & 2  & 1.00      & 1.00   & 1.00     \\
                                                                             & two-hop   & 0     & 2  & 1.00      & 1.00   & 1.00     \\ \hline
\multirow{2}{*}{\begin{tabular}[c]{@{}c@{}}TC3 \\ (TRACE)\end{tabular}}        & one-hop   & 173   & 1  & 1.00      & 1.00   & 1.00     \\
                                                                             & two-hop   & 269   & 1  & 1.00      & 1.00   & 1.00     \\ \hline
\multirow{2}{*}{\begin{tabular}[c]{@{}c@{}}TC3 \\ (THEIA)\end{tabular}}        & \textbf{one-hop}   & \textbf{0}     & \textbf{5}  & \textbf{1.00}      & \textbf{1.00}   & \textbf{1.00}     \\
                                                                             & two-hop   & 21.5K & 2  & 0.50      & 1.00   & 0.67     \\ \hline
\multirow{2}{*}{\begin{tabular}[c]{@{}c@{}}OpTC \\ (201)\end{tabular}}         & one-hop   & 0     & 7  & 1.00      & 0.88   & 0.94     \\
                                                                             & two-hop   & 3     & 6  & 0.95      & 0.90   & 0.92     \\ \hline
\multirow{2}{*}{\begin{tabular}[c]{@{}c@{}}OpTC \\ (501)\end{tabular}}         & one-hop   & 0     & 0  & 1.00      & 1.00   & 1.00     \\
                                                                             & two-hop   & 2     & 0  & 0.99      & 1.00   & 1.00     \\ \hline
\multirow{2}{*}{\begin{tabular}[c]{@{}c@{}}OpTC \\ (51)\end{tabular}}          & one-hop   & 17    & 39 & 0.89      & 0.77   & 0.82     \\
                                                                             & two-hop   & 26    & 29 & 0.84      & 0.83   & 0.84     \\ \hline
\multirow{2}{*}{\begin{tabular}[c]{@{}c@{}}NODLINK\\  (Ubuntu)\end{tabular}} & one-hop   & 1     & 0  & 0.95      & 1.00   & 0.97     \\
                                                                             & two-hop   & 0     & 0  & 1.00      & 1.00   & 1.00     \\ \hline
\multirow{2}{*}{\begin{tabular}[c]{@{}c@{}}NODLINK\\  (WS 12)\end{tabular}}  & one-hop   & 13    & 3  & 0.74      & 0.93   & 0.82     \\
                                                                             & two-hop   & 13    & 3  & 0.74      & 0.93   & 0.82     \\ \hline
\multirow{2}{*}{\begin{tabular}[c]{@{}c@{}}NODLINK\\  (W 10)\end{tabular}}   & one-hop   & 9     & 1  & 0.95      & 0.99   & 0.97     \\
                                                                             & two-hop   & 0     & 1  & 1.00      & 0.99   & 1.00     \\ \hline
\end{tabular}%
}
\end{table}
\section{Feature Selection Analysis}
\label{feature-selection-appx}
As part of our feature selection process, we evaluated two additional temporal features: Lifespan, defined as the duration between a node's first and last observed actions, and Cumulative Active Time, defined as the total time between consecutive actions with gaps under one second. To assess their effectiveness, we created three system variants: our proposed system, one with the Lifespan feature (\textit{With Lifespan}), and one with the Cumulative Active Time feature (\textit{With CumActive}). OCR-APT uses two core behavioral features---normalized action frequency and idle period statistics---selected for their ability to generalize across diverse hosts.

\begin{table}[t]
\centering
\caption{Impact of Lifespan and Cumulative Active Time (CumActive) Features on Detection Performance across datasets.}
\label{feature-selection-table}
\resizebox{\columnwidth}{!}{%
\begin{tabular}{ccccccc}
\hline
Dataset                                                                  & Version          & FP          & FN         & Precision     & Recall        & F1-Score      \\ \hline
\multirow{3}{*}{\begin{tabular}[c]{@{}c@{}}TC3 \\ (CADETS)\end{tabular}} & With Lifespan    & 0           & 1          & 1.00          & 1.00          & 1.00          \\
                                                                         & With CumActive   & 18          & 1          & 1.00          & 1.00          & 1.00          \\
                                                                         & OCR-APT          & 0           & 2          & 1.00          & 1.00          & 1.00          \\ \hline
\multirow{3}{*}{\begin{tabular}[c]{@{}c@{}}TC3 \\ (TRACE)\end{tabular}}  & With Lifespan    & 38          & 0          & 1.00          & 1.00          & 1.00          \\
                                                                         & With CumActive   & 180         & 0          & 1.00          & 1.00          & 1.00          \\
                                                                         & OCR-APT          & 173         & 1          & 1.00          & 1.00          & 1.00          \\ \hline
\multirow{3}{*}{\begin{tabular}[c]{@{}c@{}}TC3 \\ (THEIA)\end{tabular}}     & With Lifespan & 137 & 5  & 0.99 & 1.00 & 1.00 \\
                                                                         & With CumActive   & 153         & 5          & 0.99          & 1.00          & 1.00          \\
                                                                         & OCR-APT          & 0           & 5          & 1.00          & 1.00          & 1.00          \\ \hline
\multirow{3}{*}{\begin{tabular}[c]{@{}c@{}}OpTC \\ (201)\end{tabular}}   & With Lifespan    & 57          & 6          & 0.49          & 0.90          & 0.63          \\
                                                                         & With CumActive   & 55          & 16         & 0.44          & 0.73          & 0.55          \\
                                                                         & \textbf{OCR-APT} & \textbf{0}  & \textbf{7} & \textbf{1.00} & \textbf{0.88} & \textbf{0.94} \\ \hline
\multirow{3}{*}{\begin{tabular}[c]{@{}c@{}}OpTC \\ (501)\end{tabular}}   & With Lifespan    & 4           & 0          & 0.99          & 1.00          & 0.99          \\
                                                                         & With CumActive   & 1           & 0          & 1.00          & 1.00          & 1.00          \\
                                                                         & OCR-APT          & 0           & 0          & 1.00          & 1.00          & 1.00          \\ \hline
\multirow{3}{*}{\begin{tabular}[c]{@{}c@{}}OpTC \\ (51)\end{tabular}}    & With Lifespan    & 20          & 36         & 0.87          & 0.79          & 0.83          \\
                                                                         & With CumActive   & 17          & 39         & 0.89          & 0.77          & 0.82          \\
                                                                         & OCR-APT          & 17          & 39         & 0.89          & 0.77          & 0.82          \\ \hline
\multirow{3}{*}{\begin{tabular}[c]{@{}c@{}}NODLINK\\ (Ubuntu)\end{tabular}} & With Lifespan & 0   & 0  & 1.00 & 1.00 & 1.00 \\
                                                                         & With CumActive   & 23          & 0          & 0.44          & 1.00          & 0.61          \\
                                                                         & OCR-APT          & 1           & 0          & 0.95          & 1.00          & 0.97          \\ \hline
\multirow{3}{*}{\begin{tabular}[c]{@{}c@{}}NODLINK\\  (WS 12)\end{tabular}} & With Lifespan & 78  & 40 & 0.00 & 0.00 & 0.00 \\
                                                                         & With CumActive   & 54          & 40         & 0.00          & 0.00          & 0.00          \\
                                                                         & \textbf{OCR-APT} & \textbf{13} & \textbf{3} & \textbf{0.74} & \textbf{0.93} & \textbf{0.82} \\ \hline
\multirow{3}{*}{\begin{tabular}[c]{@{}c@{}}NODLINK\\  (W 10)\end{tabular}}  & With Lifespan & 0   & 1  & 1.00 & 0.99 & 1.00 \\
                                                                         & With CumActive   & 3           & 1          & 0.98          & 0.99          & 0.99          \\
                                                                         & OCR-APT          & 9           & 1          & 0.95          & 0.99          & 0.97      \\ \hline   
\end{tabular}%
}
\end{table}

As shown in Table~\ref{feature-selection-table}, the Lifespan feature yielded minor improvements on a few hosts. On OpTC 51, it slightly boosted recall compared to OCR-APT, and on Ubuntu and W10, it slightly improved precision by eliminating false positives. However, its performance dropped sharply on other hosts. On WS12, it led to complete failure---F1 score fell to zero due to missed detections (\textit{$TP = 0$}). Similarly, on OpTC 201, Lifespan caused a substantial drop in precision (from 1.00 to 0.49), severely degrading the F1 score.

The Cumulative Active Time feature showed a modest benefit only on W10, where it slightly reduced false positives compared to OCR-APT. On all other hosts, however, it either did not improve performance or led to degradation. On WS12, its inclusion once again led to detection failure, replicating the poor performance observed with Lifespan on this host. On OpTC 201, it sharply reduced precision (to 0.44), and on Ubuntu, it introduced a large number of false positives (23, compared to just one in OCR-APT).

Overall, although both features offered marginal gains on a few hosts, their lack of stability and the significant performance drops on others led us to exclude them from the final system.
We also excluded features such as \textit{most active hours} due to limited generalizability and potential dataset bias. Simulated datasets may not reflect real-world attacker behavior, as adversaries can evade detection by operating during typical business hours---periods that are increasingly difficult to define due to flexible work schedules and remote access.
Our findings validate the selected feature set, though a broader exploration of alternative temporal features remains an open direction for future research.

\end{document}